\pdfoutput=1
\documentclass[a4paper,fleqn,usenatbib,useAMS]{mnras}

\usepackage[usenames, dvipsnames]{color}
\usepackage{subcaption}
\usepackage{graphicx}
\usepackage{amsmath}
\usepackage{booktabs}

\DeclareMathOperator{\Evid}{\mathcal{Z}}
\DeclareMathOperator{\Like}{\mathcal{L}}
\DeclareMathOperator{\Model}{\mathcal{M}}
\DeclareMathOperator{\Prob}{\mathrm{Pr}}
\DeclareMathOperator{\Nlive}{\mathit{N}_{\mathrm{live}}}
\DeclareMathOperator{\Nrep}{\mathit{N}_\mathrm{rep}}
\DeclareMathOperator{\PORGR}{\mathit{P}^\mathrm{modGR}_\mathrm{GR}}
\newcommand{\codeF}[1]{\textsc{#1}}

\usepackage[T1]{fontenc}
\usepackage{ae,aecompl}

\usepackage{newtxtext,newtxmath}

\title[Towards a framework for testing GR with EMRI observations]{Towards a framework for testing general relativity with extreme-mass-ratio-inspiral observations}

\author[A. J. K. Chua et al.]{A. J. K. Chua,$^{1,2}$\thanks{Contact e-mail: \href{mailto:alvin.j.chua@jpl.nasa.gov}{alvin.j.chua@jpl.nasa.gov}}
S. Hee,$^{3,4}$
W. J. Handley,$^{3,4,5}$
E. Higson,$^{3,4}$
C. J. Moore,$^{6,7}$
\newauthor
J. R. Gair,$^{8}$
M. P. Hobson$^{3}$ and
A. N. Lasenby$^{3,4}$
\\
\\
$^{1}$ Jet Propulsion Laboratory, California Institute of Technology, 4800 Oak Grove Drive, Pasadena, CA 91109, USA \\
$^{2}$ Institute of Astronomy, Madingley Road, Cambridge, CB3 0HA, UK \\
$^{3}$ Astrophysics Group, Battcock Centre, Cavendish Laboratory, JJ Thomson Avenue, Cambridge, CB3 0HE, UK \\
$^{4}$ Kavli Institute for Cosmology Cambridge, Madingley Road, Cambridge, CB3 0HA, UK \\
$^{5}$ Gonville and Caius College, Trinity Street, Cambridge, CB2 1TA, UK \\
$^{6}$ Centro de Astrof{\'{i}}sica e Gravita{\c{c}}{\~{a}}o -- CENTRA, Departamento de F{\'{i}}sica, Instituto Superior T{\'{e}}cnico -- IST, Universidade de Lisboa -- UL,\\ Av. Rovisco Pais 1, 1049-001 Lisboa, Portugal \\
$^{7}$ Department of Applied Mathematics and Theoretical Physics, Centre for Mathematical Sciences, Wilberforce Road, Cambridge, CB3 0WA, UK \\
$^{8}$ School of Mathematics, University of Edinburgh, King's Buildings, Edinburgh, EH9 3JZ, UK}

\date{Last updated DD Month 2018; in original form DD Month 2018}

\pubyear{2018}

\begin{document}

\label{firstpage}
\pagerange{\pageref{firstpage}--\pageref{lastpage}}
\maketitle

\begin{abstract} 
Extreme-mass-ratio-inspiral observations from future space-based gravitational-wave detectors such as LISA will enable strong-field tests of general relativity with unprecedented precision, but at prohibitive computational cost if existing statistical techniques are used. In one such test that is currently employed for LIGO black-hole binary mergers, generic deviations from relativity are represented by $N$ deformation parameters in a generalised waveform model; the Bayesian evidence for each of its $2^N$ combinatorial submodels is then combined into a posterior odds ratio for modified gravity over relativity in a null-hypothesis test. We adapt and apply this test to a generalised model for extreme-mass-ratio inspirals constructed on deformed black-hole spacetimes, and focus our investigation on how computational efficiency can be increased through an evidence-free method of model selection. This method is akin to the algorithm known as product-space Markov chain Monte Carlo, but uses nested sampling and improved error estimates from a rethreading technique. We perform benchmarking and robustness checks for the method, and find order-of-magnitude computational gains over regular nested sampling in the case of synthetic data generated from the null model.
\end{abstract}

\begin{keywords}
gravitational waves -- methods: data analysis -- methods: statistical
\end{keywords}


 
\section{Introduction}
\label{sec:intro}

The spate of gravitational-wave (GW) sources found by Advanced LIGO in its first two observing runs \citep{PhysRevX.6.041015,PhysRevLett.118.221101,PhysRevLett.119.141101,PhysRevLett.119.161101,2017ApJ...851L..35A} have opened up a new branch of multimessenger astronomy---one that extends its reach beyond electromagnetic radiation for the first time, and into the gravitational sector. Traditional electromagnetic telescopes will work together with a ground-based GW detector network \citep{1742-6596-610-1-012012}, pulsar timing arrays \citep{0034-4885-78-12-124901}, and future space-based GW detectors \citep{2017arXiv170200786A} to discover and study a broad spectrum of sources that are astrophysical or cosmological in origin.

Astronomy with GW observations will also improve our understanding of gravitation and fundamental physics, by granting access to unprecedented tests of general relativity (GR) and its alternatives in the dynamical strong-field regime \citep{Yunes2013,Gair2013}. Several such tests have been performed on data from Advanced LIGO's first observing run, with no evidence for deviation from GR to date \citep{PhysRevX.6.041015,PhysRevLett.116.221101,PhysRevD.94.084002,2018arXiv180210194T,PhysRevLett.120.031104}; these include various GR consistency checks for measured signals, as well as the placing of constraints on non-tensorial GW polarisations in generic metric theories, and on the graviton Compton wavelength in massive-gravity theories.

One particular test focuses on the late-stage phase evolution of GW signals from black-hole binary mergers, which encode several strong-field effects that are not observable for binary pulsars. This approach is a particular implementation of the parametrised post-Einsteinian framework \citep{PhysRevD.80.122003}, in which model-independent deviations from GR are parametrised as deformations to the post-Newtonian (PN) and phenomenological phase parameters in a GR-based inspiral--merger--ringdown waveform model \citep{PhysRevD.93.044007}. Any phase deformations for a given signal are then constrained through comparison with the generalised model, and a Bayesian model selection framework \citep{PhysRevD.85.082003,PhysRevD.89.082001} is used to perform a null-hypothesis test of GR.

Merging binary systems have provided the only GW signals detected so far, and are expected to be ubiquitous across the bandwidths of current and future interferometric detectors. They will be an important type of source for ESA's space-based detector LISA \citep{2017arXiv170200786A}, in the form of massive-black-hole (MBH) binary mergers and the extreme-mass-ratio inspirals (EMRIs) of stellar-origin compact objects into MBHs within galactic nuclei \citep{0264-9381-29-12-124016}. EMRIs in particular have the potential to facilitate stringent tests of GR \citep{PhysRevD.75.042003,Gair2013}; this is because the phase of the GW signal can be precisely tracked over the large number of observable orbits spent by the compact object in the strong field of the central black hole.

Model-independent parametric tests of GR with EMRIs may be performed using an EMRI analogue of the generalised waveform model for comparable-mass black-hole binary mergers. One such model \citep{PhysRevD.84.064016} introduces deformations at the level of the background metric around the central MBH, by constructing analytic EMRI waveforms \citep{PhysRevD.69.082005} on a bumpy black-hole spacetime that retains an approximately conserved energy, angular momentum and Carter constant along each geodesic orbit \citep{PhysRevD.83.104027}. This model shows the potential of EMRIs for testing the Kerr solution in GR, as it can place much tighter constraints on the metric deformations than current X-ray observations \citep{PhysRevD.92.024039,0264-9381-34-19-195009}.

However, the prospect of doing precision science with EMRIs is accompanied by a high degree of technical difficulty. The length and complexity of EMRI signals leads to computationally expensive models and a highly multimodal likelihood surface in Bayesian inference problems, which exacerbates the already significant challenge of evaluating the evidence for model selection. Modern algorithms such as nested sampling can explore comparable-mass merger likelihoods efficiently \citep{0264-9381-26-21-215003,PhysRevD.81.062003} and are used to compute Bayes factors in the LIGO framework \citep{PhysRevLett.116.221101}, but improved techniques are required to adapt these algorithms for tests of GR with EMRIs.

In this paper, we take a first step towards developing a framework for testing GR with EMRI observations. The generalised EMRI waveform model described above is trialled in a Bayesian model selection framework based on the LIGO tests, using a nested-sampling algorithm that is tailored for high-dimensional and multimodal likelihoods \citep{Handley2015,Handley2015b}. We assess the viability of a product-space method \citep{Hee2015} in accelerating the convergence of nested sampling on EMRI likelihoods, and apply a rethreading technique \citep{higson2017} that provides error estimates on the Bayes factors obtained through this method.

As the generalised EMRI model is too computationally unwieldy for method development purposes, we first present a proof of principle on a toy waveform model that mirrors several of its key qualitative features. The rethreading technique is empirically validated, and is used to compare the errors attained by the product-space method and regular nested sampling, as a function of total likelihood evaluations. We find that for the toy model likelihood, the product-space method reduces by an order of magnitude the number of likelihood calls taken to reach a standard deviation of 5\% on the final GR/non-GR Bayes factor.

A similar but scaled-down analysis is then performed for the generalised model, with sampling restricted to a subset of EMRI parameters. Due to the greater complexity of the likelihood in this case, sampling bias is more likely to occur in shorter nested-sampling runs; we briefly demonstrate how a population of such runs may be combined to reduce this bias and ensure faster convergence in practice. Results for this more realistic EMRI likelihood show that the product-space method still attains better precision than regular nested sampling at the same level of computational cost, and indicate an even greater improvement in terms of cost to reach 5\% error (potentially reducing the required number of likelihood calls by two orders of magnitude).

In Section \ref{sec:models}, we give an overview of the generalised EMRI waveform model and define a toy surrogate that shares some of its relevant features. The statistical framework for testing GR with this model is set out in Section \ref{sec:framework}; we briefly summarise the key components of the LIGO test infrastructure, and describe the product-space method and rethreading technique that are used to adapt it for EMRI tests in this work. We then investigate in Section \ref{sec:toyresults} the viability of our proposed framework through a mock test of GR with the toy model, before extending our analysis to the generalised model in Section~\ref{sec:results}. Throughout this paper, we adopt geometrised units such that $c=G=1$. Greek (spacetime) indices run from 0 to 3, while Latin (space) indices run from 1 to 3. The base-10 logarithm is denoted by $\lg$, while the natural logarithm is denoted by $\ln$.

\section{Waveform models}
\label{sec:models}

The generalised model considered in this work is based on the analytic kludge (AK) formalism of \cite{PhysRevD.69.082005}, which we summarise in Section \ref{sec:AK}. This GR-based EMRI model is the most computationally efficient one available, and as such has been widely used in scoping out data analysis for space-based GW detectors; it is qualitatively resemblant to more accurate EMRI models, and its quantitative fidelity may also be improved through frequency corrections \citep{0264-9381-32-23-232002,PhysRevD.96.044005}. In Section \ref{sec:bAK}, we describe the construction by \cite{PhysRevD.84.064016} of AK waveforms on a family of generic modified-gravity black-hole spacetimes. These spacetimes are parametrised by metric deformations (or ``bumps'') of different sizes, which show up in the resultant ``bumpy AK'' (bAK) model as perturbations to the phase evolution at different orders. The sinusoidal toy model we introduce in Section \ref{sec:toy} is designed to mimic the parameter dependence of the bAK phase evolution, at a fraction of the computational cost.

\subsection{Analytic kludge model}
\label{sec:AK}

In the AK model, the instantaneous orbit of the compact object around the central black hole is approximated as Newtonian; the orbital parameters of this Keplerian ellipse are evolved over time with mixed-order PN expressions to simulate relativistic effects such as radiation reaction (inspiralling and circularising) and orbital precession (apsidal and Lense--Thirring). The waveform is then generated along the orbital trajectory using the \cite{PhysRev.131.435} mode-sum approximation.

The inertia tensor for an EMRI with masses $(\mu,M\gg\mu)$ is given by $I^{ij}(t)=\mu x^{i}(t)x^{j}(t)$, where $\mathbf{x}$ is the position of the compact object relative to the central black hole. For an instantaneous Newtonian orbit with orbital frequency $\nu$, the second time derivative of $I^{ij}$ may be decomposed into $n$-harmonics of $\nu$ as $\ddot{I}^{ij}=\sum_{n}\ddot{I}_{n}^{ij}$. The three independent components of $\ddot{I}_{n}^{ij}$ are given by
\begin{equation}\label{eq:itensor}
\ddot{I}_{n}^{11}=a_{n}+c_{n},\quad\ddot{I}_{n}^{12}=b_{n},\quad\ddot{I}_{n}^{22}=c_{n}-a_{n},
\end{equation}
with \citep{PhysRev.131.435}
\begin{align}
a_n=&-n{\mathcal{A}}\bigg[J_{n-2}(ne)-2eJ_{n-1}(ne)+\frac{2}{n}J_n(ne)+2eJ_{n+1}(ne)\nonumber\\&-J_{n+2}(ne)\bigg]\cos{(n\Phi)},\\
b_n=&-n{\cal{A}}\left(1-e^{2}\right)^{1/2}[J_{n-2}(ne)-2J_n(ne)+J_{n+2}(ne)]\nonumber\\&\times\sin{(n\Phi)},\\
c_n=&\,2{\cal{A}}J_n(ne)\cos(n\Phi),\\
\mathcal{A}=&\,\mu\tilde{\nu}^{2/3},
\end{align}
where the $J_{n}$ are Bessel functions of the first kind, $e$ is the orbital eccentricity, $\Phi(t)$ is the mean anomaly (i.e. $\dot{\Phi}=2\pi\nu$), and $\tilde{\nu}:=2\pi\nu M$ is the dimensionless orbital angular frequency.

At the detector location, it is convenient to work in an orthonormal coordinate frame $\{\hat{\mathbf{p}},\hat{\mathbf{q}},\hat{\mathbf{r}}\}$ with
\begin{equation}
\hat{\mathbf{p}}=\frac{\hat{\mathbf{r}}\times\hat{\mathbf{L}}}{\left|\hat{\mathbf{r}}\times\hat{\mathbf{L}}\right|},\quad\hat{\mathbf{q}}=\hat{\mathbf{p}}\times\hat{\mathbf{r}},
\end{equation}
where $\hat{\mathbf{r}}$ points from detector to source and $\mathbf{L}$ is the orbital angular momentum of the binary. In the transverse--traceless gauge, the leading-order gravitational radiation at the detector due to a source at luminosity distance $D$ is given by \citep{1973grav.book.....M}
\begin{equation}
h_{ij}=\frac{2}{D}\left(P_{ik}P_{jl}-\frac{1}{2}P_{ij}P_{kl}\right)\ddot{I}^{kl},\quad h^{+,\times}=\frac{1}{2}h^{ij}H_{ij}^{+,\times},
\end{equation}
with the polarisation and projection tensors
\begin{equation}\label{eq:pptensors}
H_{ij}^{+}=\hat{p}_i\hat{p}_j-\hat{q}_i\hat{q}_j,\quad H_{ij}^{\times}=\hat{p}_i\hat{q}_j+\hat{q}_i\hat{p}_j,\quad P_{ij}=\delta_{ij}-\hat{r}_i\hat{r}_j,
\end{equation}
where $\delta_{ij}$ is the Kronecker delta.

From \eqref{eq:itensor}--\eqref{eq:pptensors}, the two GW polarisation amplitudes $h^{+,\times}$ for an EMRI may be written in terms of the Peters--Mathews harmonic decomposition as \citep{PhysRevD.69.082005}
\begin{align}
&h^{+,\times}=\frac{1}{D}\sum_n A^{+,\times}_n,\\
&A^+_n=C^+_aa_n+C^+_bb_n+C^+_cc_n,\\
&A^\times_n=C^\times_aa_n+C^\times_bb_n,
\end{align}
where the coefficients $C^{+,\times}_{a,b,c}$ depend on the angular configuration $(\lambda,\tilde{\gamma},\alpha,\theta_K,\phi_K,\theta_S,\phi_S)$ of the EMRI: the inclination $\lambda$ between $\mathbf{L}$ and the black-hole spin $\mathbf{S}$; the azimuth $\tilde{\gamma}$ of periapsis in the orbital plane (relative to $\mathbf{L}\times\mathbf{S}$); the azimuth $\alpha$ of $\mathbf{L}$ projected onto the spin-equatorial plane; the orientation $(\theta_K,\phi_K)$ of $\mathbf{S}$; and the sky location $(\theta_S,\phi_S)$. The first two angles are intrinsic to the source, while the rest are defined relative to a fixed ecliptic-based coordinate system \citep{PhysRevD.57.7089}. Explicit expressions for the coefficients are given by Eqs. (10) and (18)--(25) in \cite{PhysRevD.69.082005}.

In a relativistic EMRI, the frequency $\nu$, eccentricity $e$ and inclination $\lambda$ change over time due to radiation reaction, while the angles $(\tilde{\gamma},\alpha)$ precess as well. The AK model evolves $(\nu,e)$ with 3.5PN expressions, while approximating $\lambda$ as constant due to its slow evolution over the full inspiral \citep{PhysRevD.61.084004}. The fluxes $(\dot{\nu},\dot{e})$ are given by \citep{1992MNRAS.254..146J}
\begin{align}\label{eq:fluxes}
&\dot{\nu}=\frac{96}{10\pi}\frac{\eta\tilde{\nu}^{11/3}}{M^2(1-e^2)^{7/2}}\left(1+\frac{73e^2}{24}+\frac{37e^4}{96}\right)+\mathcal{O}(\tilde{\nu}^{13/3}),\\
&\dot{e}=-\frac{304}{15}\frac{\eta\tilde{\nu}^{8/3}e}{M(1-e^2)^{5/2}}\left(1+\frac{121e^2}{304}\right)+\mathcal{O}(\tilde{\nu}^{10/3}),
\end{align}
where $\eta:=\mu/M$ is the mass ratio, and only the leading 2.5PN terms are explicitly presented here. (See Eqs. (28) and (30) in \cite{PhysRevD.69.082005} for the full expressions.)

The angular rates $(\dot{\tilde{\gamma}},\dot{\alpha})$ determine the precession of periapsis in the orbital plane, $\dot{\tilde{\gamma}}+\dot{\alpha}$, and the Lense--Thirring precession of the orbital plane around the black-hole spin axis, $\dot{\alpha}$. They are given by \citep{1992MNRAS.254..146J,PhysRevD.12.329}
\begin{align}
&\dot{\tilde{\gamma}}=3\frac{\tilde{\nu}^{5/3}}{M(1-e^2)}+\mathcal{O}(\tilde{\nu}^2),\label{eq:gammadot}\\
&\dot{\alpha}=2\frac{\tilde{a}\tilde{\nu}^2}{M(1-e^2)^{3/2}},\label{eq:alphadot}
\end{align}
where $\tilde{a}:=|\mathbf{S}|/M^2$ is the dimensionless spin angular momentum. (See Eq. (29) in \cite{PhysRevD.69.082005} for the full version of \eqref{eq:gammadot}.)

In the original AK model, the two precession rates and the orbital frequency can differ significantly from their actual values in a relativistic EMRI. Following \cite{0264-9381-32-23-232002}, we correct the starting angular rates $(\dot{\Phi}_0,\dot{\tilde{\gamma}}_0,\dot{\alpha}_0)$ through a parameter-space map $(M,\tilde{a},\nu)\mapsto(M',\tilde{a}',\nu')$ such that
\begin{align}
&\dot{\Phi}_0(M',\tilde{a}',\nu')=\omega_r(M,\tilde{a},\nu),\\
&\dot{\tilde{\gamma}}_0(M',\tilde{a}',\nu')=\omega_\theta(M,\tilde{a},\nu)-\omega_r(M,\tilde{a},\nu),\\
&\dot{\alpha}_0(M',\tilde{a}',\nu')=\omega_\phi(M,\tilde{a},\nu)-\omega_\theta(M,\tilde{a},\nu),
\end{align}
where $\omega_{r,\theta,\phi}$ are the fundamental frequencies of radial, polar and azimuthal motion \citep{0264-9381-19-10-314} on the starting Kerr geodesic. This map does not eradicate the accumulated phase error of AK waveforms over the full inspiral, but has negligible computational cost and greatly improves the quantitative accuracy over short times.

An EMRI has 14 degrees of freedom in the AK model, which neglects the spin of the compact object; the parameters of the model are chosen to decouple the seven source-intrinsic degrees of freedom from the observer-dependent extrinsic ones. We define the set of dimensionless AK parameters as
\begin{align}
&\Theta_\mathrm{AK}=\Theta_\mathrm{int}\cup\Theta_\mathrm{ext},\\
&\Theta_\mathrm{int}=\left\{\lg{\left(\frac{\mu}{M_\odot}\right)},\lg{\left(\frac{M}{M_\odot}\right)},\tilde{a},e_0,\cos{\lambda},\Phi_0,\tilde{\gamma}_0\right\},\label{eq:intrinsic}\\
&\Theta_\mathrm{ext}=\left\{\tilde{\nu}_0,\alpha_0,\cos{\theta_K},\phi_K,\cos{\theta_S},\phi_S,\lg{\left(\frac{D}{\mathrm{Gpc}}\right)}\right\},
\end{align}
where the subscript zero denotes the value taken by a quantity at the arbitrary starting time $t=0$.

\subsection{Bumpy analytic kludge model}
\label{sec:bAK}

The family of generically deformed black-hole spacetimes derived separately by \cite{Benenti1979} and \cite{PhysRevD.83.104027} provides a model-independent setting in which to construct modified-gravity EMRI waveforms for testing the Kerr metric solution in GR. These bumpy black holes are not required to satisfy the Einstein equations (as done in \cite{PhysRevD.69.124022,PhysRevD.81.024030}), but are Kerr-like through their stationarity, axisymmetry, and admission of an approximate second-rank Killing tensor. They are also more general than the modified Kerr spacetimes considered in other proposed EMRI tests \citep{0264-9381-23-12-013,PhysRevD.75.042003}, as they allow for mass-moment deformations beyond quadrupole order.

In Boyer--Lindquist coordinates, the components of the Kerr metric around a black hole with mass $M$ and spin angular momentum $a=\tilde{a}M$ are given by \citep{1973grav.book.....M}
\begin{align}
&g^\mathrm{K}_{tt}=-\left(1-\frac{2Mr}{\Sigma}\right),\quad g^\mathrm{K}_{t\phi}=-\frac{2Mar\sin^2{\theta}}{\Sigma},\quad g^\mathrm{K}_{rr}=\frac{\Sigma}{\Delta},\nonumber\\
&g^\mathrm{K}_{\theta\theta}=\Sigma,\quad g^\mathrm{K}_{\phi\phi}=\left(r^2+a^2+\frac{2Ma^2r\sin^2{\theta}}{\Sigma}\right)\sin^2{\theta},
\end{align}
where $\Sigma:=r^2+a^2\cos^2{\theta}$ and $\Delta:=r^2-2Mr+a^2$. The Kerr metric is stationary and axisymmetric, and hence may be cast in Lewis--Papapetrou form via the partial coordinate transformation
\begin{equation}\label{eq:transform}
(\rho,z)=\left(\sqrt{\Delta}\sin{\theta},(r-M)\sin{\theta}\right),
\end{equation}
with the temporal and azimuthal coordinates $(t,\phi)$ left unchanged.

\cite{PhysRevD.83.104027} consider a linear deformation of the Kerr metric in Lewis--Papapetrou form, and apply the inverse of the transformation \eqref{eq:transform} to obtain its components in Boyer--Lindquist-like coordinates $(r,\theta)$; this ensures that the deformed metric remains stationary and axisymmetric, i.e. it admits temporal and azimuthal Killing vectors $t^\mu$ and $l^\mu$ such that
\begin{equation}
\nabla_{(\mu}t_{\nu )}=\nabla_{(\mu}l_{\nu )}=0.
\end{equation}
The deformed metric components may then be written as
\begin{equation}
g_{\mu\nu}=g^\mathrm{K}_{\mu\nu}+\epsilon h_{\mu\nu},
\end{equation}
where $\epsilon\ll1$ is a bookkeeping parameter for the metric deformation $h_{\mu\nu}$. Other Kerr-like properties are included by requiring the second-rank tensor $\xi_{\mu\nu}=\Delta t_{(\mu}l_{\nu)}+r^2g_{\mu\nu}$ to approximately satisfy the Killing tensor equation, i.e.
\begin{equation}
\nabla_{(\lambda}\xi_{\mu\nu )}=\mathcal{O}(\epsilon^2),
\end{equation}
and by requiring $|h_{\mu\nu}|=\mathcal{O}(1/r^2)$ as $M/r\to0$ (such that the deformed metric retains asymptotic flatness, along with its original mass and spin angular momentum).

With the above constraints, the only nonzero components of $h_{\mu\nu}$ are $h_{tt}$, $h_{t\phi}$, $h_{rr}$ and $h_{\phi\phi}$, which depend on the black-hole parameters $(M,\tilde{a})$ and three arbitrary radial functions $\gamma_i,i\in\{1,3,4\}$. By representing the nonzero components and radial functions as power series in $M/r$, \cite{PhysRevD.84.064016} obtain expressions for $h_{\mu\nu,n},2\leq n\leq5$ in terms of $\gamma_{i,n}$, where these quantities are the coefficients of the $(M/r)^n$ terms in the corresponding series. If the inclination angle of a geodesic orbit is approximated as constant (as done in the AK model), a metric deformation that is purely\footnote{For simplicity, the allowed deformations in the bAK model are restricted to ``pure'' bumps and their linear combinations, e.g. a $\mathcal{B}_4$ bump is not the most general 4th-order deformation, but defined as one for which the $\mathcal{B}_{2,3,5}$ coefficients are all zero. While the model might not fully represent deviations from GR at sub-leading order, it is an adequate surrogate in the present work.} $n$th-order in $M/r$ turns out to be fully specified by a set of three coefficients, $\mathcal{B}_n:=\{\gamma_{1,n},\gamma_{4,n},\gamma_{3,n+1}\}$; we refer to such a deformation as a $\mathcal{B}_n$ bump.

The AK formalism is then used to construct EMRI waveforms on the bumpy black-hole spacetime, which is parametrised by the set of coefficients $\bigcup_n\mathcal{B}_n,2\leq n\leq5$ in addition to $(M,\tilde{a})$. Both the long-timescale radiation reaction fluxes and the short-timescale precession rates of the EMRI are altered by the $\mathcal{B}_n$ coefficients, since they perturb the three first integrals of motion along a timelike geodesic with four-velocity $u^\mu$: the energy $E=t_\mu u^\mu$ and orbital angular momentum $L_z=l_\mu u^\mu$, which remain conserved, and the analogue of the Carter constant $Q=\xi_{\mu\nu}u^\mu u^\nu$, which is conserved at linear order in the metric deformation.

By matching the turning points of motion for an instantaneous geodesic orbit $(E,L_z,Q)$ with those of a precessing Keplerian orbit $(\nu,e,\lambda)$ in the AK model, \cite{PhysRevD.84.064016} compute the leading-order corrections caused by the $\mathcal{B}_n$ deformations to the fluxes and angular rates \eqref{eq:fluxes}--\eqref{eq:alphadot}. Each set of corrections at $n$th-order in $M/r$ depends only on a single linear combination $\epsilon_n:=\gamma_{1,n}+2\gamma_{4,n}$ of the $\mathcal{B}_n$ coefficients ($\gamma_{3,n+1}$ is at sub-leading order), which effectively reduces the additional degrees of freedom in the extended model to four. The corrections are given by
\begin{align}
&\delta\dot{\nu}_n=\frac{8}{5\pi}\frac{\eta\tilde{\nu}^{(2n+9)/3}}{M^2(1-e^2)^{n+5/2}}g_{\nu,n}\epsilon_n,\label{eq:deltanudot}\\
&\delta\dot{e}_n=-\frac{16}{5}\frac{\eta\tilde{\nu}^{(2n+6)/3}}{M(1-e^2)^{n+3/2}}g_{e,n}\epsilon_n,\\
&\delta\dot{\tilde{\gamma}}_n=\frac{\tilde{\nu}^{(2n+1)/3}}{M(1-e^2)^{n-1}}g_{\tilde{\gamma},n}\epsilon_n,\\
&\delta\dot{\alpha}_n=-\frac{\tilde{\alpha}\tilde{\nu}^{(2n+2)/3}}{M(1-e^2)^{n-1/2}}g_{\alpha,n}\epsilon_n,\label{eq:deltaalphadot}
\end{align}
where explicit expressions for the eccentricity-dependent factors $g_{\cdot,n}$ are given by Eqs. (301), (302) and (335)--(346) in \cite{PhysRevD.84.064016} (with $g_{\tilde{\gamma},2}=1/2$ and $g_{\alpha,2}=1$).

\begin{figure}
\includegraphics[width=\columnwidth]{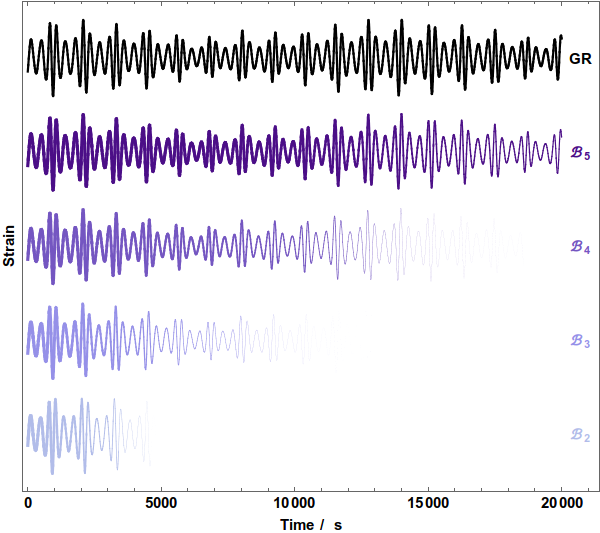}
\caption{Comparison of bAK waveforms with common source parameters and different $\mathcal{B}_n$ bumps of the same magnitude ($\epsilon_n=0.4$). These waveforms start in phase with the GR-based AK waveform by construction, but lose phase coherence (indicated by opacity) over time. Higher-order $\mathcal{B}_n$ waveforms correspond to smaller metric deformations, and thus dephase more slowly. Figure has been reproduced from \protect\cite{0264-9381-34-19-195009}.}
\label{fig:bAK}
\end{figure}

In summary, the bAK model comprises i) adding the corrections \eqref{eq:deltanudot}--\eqref{eq:deltaalphadot} to the corresponding evolution equations \eqref{eq:fluxes}--\eqref{eq:alphadot} in the AK model, and ii) extending the set of model parameters to
\begin{align}
&\Theta_\mathrm{bAK}=\Theta_\mathrm{AK}\cup\Theta_\mathcal{B},\label{eq:bAKparams}\\
&\Theta_\mathcal{B}=\{\lg{\epsilon_n}\,|\,2\leq n \leq5\},\label{eq:epsparams}
\end{align}
where we assume $\epsilon_n>0$ for simplicity. The deformation parameters $\epsilon_n$ then determine the magnitudes of the $\mathcal{B}_n$ bumps, which manifest as phase drifts in the corresponding bAK waveforms. Figure \ref{fig:bAK} illustrates how these waveforms dephase over time relative to the AK waveform (where $\epsilon_n=0$ for all $n$). Full EMRI waveforms can have up to $\sim10^5$ observable cycles, and hence might be able to constrain the leading-order bumps down to $\epsilon_n\sim10^{-7}$ \citep{0264-9381-34-19-195009}. For the sake of comparison, combined results for the LIGO events GW150914 and GW151226 in a less conservative single-parameter analysis placed no upper bound below $\sim10^{-1}$ on the dimensionless deformation parameters of the generalised black-hole binary merger model \citep{PhysRevX.6.041015}.

\subsection{Sinusoidal toy model}
\label{sec:toy}

Motivated by the phase behaviour of the bAK waveforms, we define a deformed sinusoidal model in which the phase evolution has similar dependence on a set of deformation parameters analogous to $\Theta_\mathcal{B}$. Waveforms from this toy model are several orders of magnitude faster to compute than bAK waveforms, and as such are useful for building intuition during method development. They are qualitatively equivalent to the special case of bAK waveforms from circular, equatorial EMRIs over short timescales (or alternatively, in the geodesic limit $\eta=0$).

Our toy model consists simply of a sine wave parametrised by amplitude $A$ and angular frequency $\Omega$, along with four parameters that deform the phase at different orders in time; these deformation parameters are denoted $\epsilon_n$ by way of analogy to the bAK model. The toy waveform is given by\footnote{Although full GW models require two independent waveform modes for parameter estimation, the likelihood obtained with the single toy model mode is qualitatively sufficient for the purposes of this work.}
\begin{equation}
h=A\sin{\left[\Omega\tilde{t}\left(1+\sum_{n=2}^5\epsilon_n\left(\frac{\tilde{t}}{\tau}\right)^{n-1}\right)\right]},
\end{equation}
where $\Omega$, $\tilde{t}$ and $\tau$ are all dimensionless. The timescale-like quantity $\tau$ has been introduced to control the overall strength of the phase drift, and is not treated as a parameter in the model. Henceforth we fix $\tau=2T_\mathrm{obs}$, where $T_\mathrm{obs}$ is some specified waveform duration, and define the set of toy model parameters as
\begin{equation}\label{eq:toyparams}
\Theta_\mathrm{toy}=\{A,\Omega\}\cup\Theta_\mathcal{B},
\end{equation}
where $\Theta_\mathcal{B}$ is defined as in \eqref{eq:epsparams}. 

\section{Statistical framework}
\label{sec:framework}

The framework presented in this paper is based on the model-independent and parametric test infrastructure introduced by \cite{PhysRevD.85.082003}, in which the combinatorial submodels of a generalised waveform model are used in a null-hypothesis test of GR, and a nested-sampling algorithm is employed to compute the Bayesian evidence for each submodel. An overview of the pertinent concepts is given in Sections \ref{sec:likelihood}--\ref{sec:nested}. In Section \ref{sec:product}, we describe the product-space nested-sampling method considered by \cite{Hee2015}; it provides an evidence-free computation of the Bayes factor for each submodel (with respect to the null submodel), thus mitigating the prohibitive computational difficulty of performing the test with a generalised EMRI waveform model. The estimation of Bayes factor errors in the method also requires a new rethreading technique \citep{higson2017}, which we outline in Section \ref{sec:rethreading}.

\subsection{Gravitational-wave likelihood}
\label{sec:likelihood}

In the basic matched-filtering framework for GW data analysis, data from a single interferometric detector is written as the time series $x=h+n$, where $h$ is the detector response to a passing GW and $n$ is the detector noise (typically approximated as a Gaussian and stationary random process). For a LISA-like detector with three arms, two independent signals $h_{I,II}$ may be obtained; these are related on average to the two GW polarisations $h^{+,\times}$ by
\begin{equation}
h_{I,II}=\frac{\sqrt{3}}{2}\left(F_{I,II}^{+}h^{+}+F_{I,II}^{\times}h^{\times}\right),
\end{equation}
where the antenna pattern functions $F_{I,II}^{+,\times}$ \citep{PhysRevD.49.6274} depend on the sky location and polarisation angle of the source in a detector-based coordinate system. (See Eqs. (15)--(17) in \cite{PhysRevD.69.082005} for the explicit expressions in ecliptic coordinates.) Doppler modulation of the waveform phase may also be included in $h^{+,\times}$ to account for the orbital motion of LISA.

For compactness, we write the GW signal as a complex time series $h=h_I+ih_{II}$.\footnote{This notation is also compatible with the single real waveform mode $h$ of the sinusoidal toy model.} The signal-to-noise ratio (SNR) of $h$ is given by $\rho:=\sqrt{\langle h|h\rangle}=\sqrt{\langle h_I|h_I\rangle+\langle h_{II}|h_{II}\rangle}$, with the noise-weighted inner product $\langle\cdot|\cdot\rangle$ defined as \citep{PhysRevD.49.2658}
\begin{equation}\label{eq:inner}
\langle a|b\rangle=2\int_0^\infty df\,\frac{\tilde{a}^*(f)\tilde{b}(f)+\tilde{a}(f)\tilde{b}^*(f)}{S_n(f)},
\end{equation}
where $S_n$ is the power spectral density of $n$. If the detector noise is assumed to be white (as done in Section \ref{sec:toyresults}), \eqref{eq:inner} simplifies to
\begin{equation}
\langle a|b\rangle=\frac{1}{S_n}\int_0^{T_\mathrm{obs}}dt\,a^*(t)b(t),
\end{equation}
which may be computed directly from the time-domain waveforms. In Section \ref{sec:results}, $S_n$ is given instead by an analytic approximation \citep{2013GWN.....6....4A} to the LISA noise power spectral density.\footnote{The noise model used corresponds to a down-scoped version of LISA, and thus gives more conservative results; the mission design has now been restored to an earlier configuration with higher sensitivity.} Throughout this paper, we work with waveforms that are renormalised with respect to some reference waveform $h'$ and specified amplitude $\rho'$, i.e. $h\to\rho'h/\sqrt{\langle h'|h'\rangle}$; hence it is more useful to think of ``SNR'' here as an amplitude relative to the defined norm in \eqref{eq:inner}, and not to the actual noise in the data.

Given GW data $x$ that contains a signal $h(\theta_*)$ corresponding to the model parameter values $\theta_*$, the Bayesian likelihood $\Like=\Prob(x|\theta)$ is defined as \citep{PhysRevD.49.2658}
\begin{equation}\label{eq:like}
\Like(\theta)\propto\exp{\left(-\frac{1}{2}\langle x-h(\theta)|x-h(\theta)\rangle\right)}.
\end{equation}
We consider only EMRI waveforms with $\rho\gtrsim10$ throughout this paper, which is a conservative choice that is consistent with typical values of the threshold SNR for reliable detection \citep{PhysRevD.95.103012,PhysRevD.96.044005}. In this regime, the noise term in the log-likelihood is suppressed by a factor of $\gtrsim100$ relative to the leading term, and may be neglected for the purposes of model selection. Hence, to simplify analysis in Sections \ref{sec:toyresults}--\ref{sec:results}, we assume a particular noise realisation of $n=0$ in the synthetic data (i.e. $x=h(\theta_*)$).

\subsection{Null-hypothesis test}
\label{sec:test}

From Bayes' theorem, the posterior probability of a model hypothesis $\Model$ given the data $x$ may be written as
\begin{equation}
\Prob(\Model|x)=\frac{\Evid\Pi}{\Prob(x)},
\end{equation}
where $\Evid=\Prob(x|\Model)$ and $\Pi=\Prob(\Model)$ are, respectively, the evidence and prior probability for the model. The model evidence is typically obtained by marginalising the likelihood $\Like=\Prob(x|\theta,\Model)$ over the model parameters, i.e.
\begin{equation}\label{eq:evidence}
\Evid=\int d\theta\,\Like(\theta)\pi(\theta),
\end{equation}
where $\pi=\Prob(\theta|\Model)$ is the parameter prior.

Two model hypotheses $\Model_{i,j}$ may be compared through the ratio of their posterior probabilities, which quantifies the degree of belief in one model over the other. The logarithm of this posterior odds ratio is defined as
\begin{equation}\label{eq:logPOR}
P^i_j:=\ln{\left[\frac{\Prob(\Model_i|x)}{\Prob(\Model_j|x)}\right]}=B^i_j+\ln{\left(\frac{\Pi_i}{\Pi_j}\right)},
\end{equation}
where $B^i_j:=\ln{(\Evid_i/\Evid_j)}$. The quantity $B^i_j$ is the logarithm of the Bayes factor, which is commonly used as an equivalent substitute for the posterior odds ratio through the implicit assumption $\Pi_i=\Pi_j$. A general scale for the interpretation of posterior odds ratios \citep{Jeffreys1961,doi:10.1080/01621459.1995.10476572} is given in Table \ref{tab:scale}.

\begin{table}
\begin{center}
\begin{tabular}{ll}
\hline
$P^i_j$ & Odds for $\Model_i$ over $\Model_j$ \\
\hline
$0\lesssim P^i_j\lesssim1$ & None \\
$1\lesssim P^i_j\lesssim3$ & Slight\\
$3\lesssim P^i_j\lesssim5$ & Significant\\
$5\lesssim P^i_j$ & Decisive\\
\hline
\end{tabular}
\end{center}
\caption{Scale for interpreting posterior odds ratios. Since $P^i_j=-P^j_i$, negative values of $P^i_j$ correspond to reversed model odds.}
\label{tab:scale}
\end{table}

For a generalised waveform model that extends a GR-based model $\Model_\mathrm{GR}$ through a set $\Theta_\mathcal{B}$ of $N$ deformation parameters, we may define $2^N$ submodels whose deformation parameter sets are given by the $2^N$ subsets of $\Theta_\mathcal{B}$. The submodel corresponding to the null set is $\Model_\mathrm{GR}$ itself; the remaining $2^N-1$ submodels are modified-GR models, which we denote collectively by the hypothesis $\Model_\mathrm{modGR}=\bigvee_{m\neq\mathrm{GR}}\Model_m$. Even though $\Model_\mathrm{modGR}$ is a collection of nested submodels, \cite{PhysRevD.85.082003} observe that the individual submodel evidences are logically disjoint, due to the distinct integration measure on each submodel parameter space. Hence the posterior odds ratio for $\Model_\mathrm{modGR}$ over $\Model_\mathrm{GR}$ simplifies to
\begin{align}
\frac{\Prob(\Model_\mathrm{modGR}|x)}{\Prob(\Model_\mathrm{GR}|x)}&=\frac{\sum_{m\neq\mathrm{GR}}\Prob(\Model_m|x)}{\Prob(\Model_\mathrm{GR}|x)}\nonumber\\&=\frac{1}{2^N-1}\sum_{m\neq\mathrm{GR}}\frac{\Evid_m}{\Evid_\mathrm{GR}},\label{eq:PORGR}
\end{align}
where the second equality follows from the assumptions that $\Pi_\mathrm{modGR}=\Pi_\mathrm{GR}$ and $\Pi_m=\Pi_{m'}$ for all $m,m'\neq\mathrm{GR}$.

This approach provides the basic framework for a test of GR with the bAK model, where the modified-GR hypothesis is compared against the null hypothesis (the AK model) through evaluation of the submodel Bayes factors in \eqref{eq:PORGR}. A single EMRI source is considered for the assessment of our methods, but it is straightforward to generalise \eqref{eq:PORGR} for a population of sources \citep{PhysRevD.85.082003}. In the bAK model (where $N=4$), the 16 submodels may be indexed by $0\leq m\leq15$, whose nonzero digits in binary representation specify the deformation parameters included in each submodel. Hence the AK model is denoted $\Model_0\equiv\Model_{0000}$, while the full bAK model is $\Model_{15}\equiv\Model_{1111}$. We use the convention that the binary digits of $m$ from right to left correspond to $\epsilon_n$ with $n=2,3,4,5$ respectively; for example, $\Model_{1010}$ has only two deformation parameters $\{\epsilon_3,\epsilon_5\}$, and is equivalent to the $\epsilon_2=\epsilon_4=0$ slice of the full model.

Although the Bayesian evidence \eqref{eq:evidence} has a built-in Occam penalty on model complexity, it does not fully account for the multiplicity effect \citep{Jeffreys1961,scott2010}. As a larger number of deformation parameters is considered, it becomes more likely that one particular parameter will cause the inclusive submodels to fit the data well by chance. In other words, a uniform submodel prior (i.e. $\Pi_m=\Pi_{m'}$ for all $m,m'\neq0$) might not be appropriate for the nested submodels in this framework, and more thorough prescriptions for assigning model prior probability \citep{doi:10.1093/biomet/87.4.731,consonni2013,SJOS:SJOS12145} should be investigated. However, the number of deformation parameters in the bAK model is small; furthermore, the focus of this work is the evaluation of \eqref{eq:PORGR} with improved precision, and not the validation of its accuracy. Hence we retain a uniform submodel prior for simplicity, and it follows from \eqref{eq:logPOR} and \eqref{eq:PORGR} that
\begin{equation}\label{eq:logPORGR}
\PORGR=B^\mathrm{modGR}_\mathrm{GR}=\ln{\left[\sum_{m=1}^{15}\exp{\left(B^m_0\right)}\right]}-\ln{15}.
\end{equation}

\subsection{Nested sampling}
\label{sec:nested}

For most statistical problems, \eqref{eq:evidence} admits no analytic solution and is impractical to evaluate through direct numerical integration, even over a parameter space of modest dimensionality. A wealth of alternative techniques for computing the evidence have thus been developed; these range from simple estimates that use Laplace's approximation \citep{Tierney1986} or Chib's method \citep{10.2307/2291521}, to more sophisticated sampling strategies based on concepts such as thermodynamic integration and simulated annealing \citep{10.2307/24306045,Gelman1998,neal2001annealed}. We employ in this work the nested sampling strategy introduced by \cite{Skilling2004}, which provides an accurate and computationally efficient means of simultaneously exploring the posterior surface and evaluating the evidence. Nested sampling has been shown to be suitable for GW likelihoods \citep{0264-9381-26-21-215003,PhysRevD.81.062003}, and is one of the two algorithms used to compute model evidences in LIGO tests of GR \citep{PhysRevLett.116.221101}.

In nested sampling, the multidimensional integral \eqref{eq:evidence} is written in the one-dimensional form \citep{Skilling2006}
\begin{equation}\label{eq:NSevidence}
\Evid=\int_0^1dX\,\Like(X),
\end{equation}
where the prior mass $X$ is given by
\begin{equation} 
X(\lambda)=\int_{\Like(\theta)>\lambda}d\theta\,\pi(\theta),
\end{equation}
i.e. the integral of the parameter prior over the interior of the likelihood contour $\Like=\lambda$. A set of $\Nlive$ initial points is first sampled from the prior; this set of ``live'' points is then evolved across parameter space by discarding at the $i$-th iteration the point $\theta_i$ with the lowest likelihood value $\lambda_i$, and replacing it with one drawn from the prior but within the contour $\Like_i=\lambda_i$.

\cite{Skilling2006} shows that the prior mass corresponding to each $\lambda_i$ may be approximated as
\begin{equation}\label{eq:NSpriormass}
X_i\approx\exp{\left(-\frac{i}{\Nlive}\right)},
\end{equation}
which shrinks exponentially from $X_0:=1$ to zero as the live points converge on and navigate the bulk of the posterior surface. Upon the satisfaction of suitable convergence criteria, the algorithm is truncated and the evidence \eqref{eq:NSevidence} may be approximated as
\begin{equation}\label{eq:NSevidest}
\Evid\approx\sum_{i>0}w_i\Like_i,
\end{equation}
where the weights $w_i$ are given by a Riemann sum rule, e.g. $w_i=X_{i-1}-X_i$. The set of discarded live points $\theta_i$ (often termed dead points) also serves as a set of posterior samples, through the assignment of posterior probability $p_i=w_i\Like_i/\Evid$ to each point.

The prior-mass approximation \eqref{eq:NSpriormass} introduces statistical error into the posterior probabilities $p_i$ and the evidence estimate \eqref{eq:NSevidest}, via the weights $w_i$. The uncertainty in the evidence estimate depends on the absolute error of each $w_i$, and is dominated by the Poisson variability in the number of iterations taken to reach the posterior bulk. Hence \eqref{eq:NSevidest} is log-normally distributed, and the standard deviation of $\ln{\Evid}$ in nested sampling scales with the number of live points as \citep{Skilling2006}
\begin{equation}\label{eq:logeviderror}
\sigma_{\ln{\Evid}}\propto\frac{1}{\sqrt{\Nlive}}.
\end{equation}

It is often difficult to sample from the prior under the constraint $\Like>\lambda$, since the likelihood contours $\Like=\lambda$ might in general be multimodal or degenerate. In this work, we make use of the nested-sampling implementation \codeF{PolyChord} \citep{Handley2015,Handley2015b}, which mitigates these difficulties through the incorporation of clustering and slice sampling algorithms. It also exhibits good scaling with the dimensionality of the parameter space, and in that sense is an improved successor to the widely adopted nested-sampling tool \codeF{MultiNest} \citep{Feroz2009,Feroz2013}. The two main runtime parameters in \codeF{PolyChord} that determine sampling resolution and reliability are, respectively, $\Nlive$ and $\Nrep$; the latter is the number of randomly oriented one-dimensional slices sampled at each iteration in order to decorrelate the new live point from the discarded point.

\subsection{Product-space nested sampling}
\label{sec:product}

Methods exist for obtaining Bayes factors without explicitly evaluating evidences, and these are especially useful when the number of competing models is large. The most well known is the Savage--Dickey density ratio for nested models (and generalisations thereof \citep{Verdinelli1995,marin2010,wetzels2010encompassing}), where the Bayes factor between a null model and an encompassing one is given by the ratio between the posterior and prior probabilities at the null point in the larger model space. Another class of methods uses the product-space representation \citep{10.2307/2346151} for a prespecified and indexed collection of competing models; this approach involves exploring the set of model indices and each model space simultaneously, and has been investigated in the context of Markov chain Monte Carlo (MCMC) algorithms \citep{10.2307/2346151,doi:10.1198/10618600152627924,Lodewyckx2011} and nested sampling \citep{Hee2015}.

Product-space sampling and reversible-jump MCMC \citep{doi:10.1093/biomet/82.4.711} are modern examples of transdimensional frameworks for Bayesian model selection \citep{Sisson2005}. One disadvantage of product-space methods (as compared to reversible-jump MCMC) is the required prior specification of all the competing models. This is not an issue in our framework since the bAK submodels are fully defined by the power set of $\Theta_\mathcal{B}$, and the $\mathcal{B}_n$ bumps are truncated at $n=5$ (their effects are exponentially suppressed as $n$ increases). In the product-space representation, the 16 submodels $\Model_m$ in Section \ref{sec:test} may be thought of as distinct ``slices'' of some hypermodel $\Model$; the parameter space of $\Model$ is the combination\footnote{More precisely, the hypermodel parameter space is the vector bundle of submodel spaces over the discrete set of model indices.} of all the submodel spaces, and is parametrised by
\begin{equation}
\Theta=\Theta_\mathrm{gen}\cup\{m\},
\end{equation}
where $\Theta_\mathrm{gen}$ is the parameter set for the generalised waveform model (i.e. \eqref{eq:bAKparams} or \eqref{eq:toyparams}) and $m$ is the submodel index.

The posterior probability for $m$ is given by \citep{Hee2015}
\begin{align}
\Prob(m|x,\Model)&=\int d\theta\,\Prob(\theta,m|x,\Model)\nonumber\\&=\frac{1}{\Evid_{\Model}}\int d\theta\,\Like(\theta,m)\pi(\theta,m)\nonumber\\&=\frac{\pi(m)}{\Evid_{\Model}}\int d\theta_m\,\Like(\theta_m)\pi(\theta_m|m)\nonumber\\&=\frac{\pi(m)\Evid_m}{\Evid_{\Model}},
\end{align}
where $\Evid_{\Model}$ and $\Evid_m$ are the evidences for the hypermodel and $m$th submodel respectively. To obtain the third equality, we have decomposed $\theta$ into the parameters $\theta_m$ that are included in the submodel $\Model_m$, and the parameters $\phi_m$ that are excluded; the integral $\int d\phi_m\,\pi(\phi_m)=1$ then factors out of the expression.

When sampling in the hypermodel space, it is convenient to assign a uniform prior $\pi(m)=1/16$ on the submodel index, since it is straightforward to restore the assumption $\sum_{m\neq0}\Pi_m=\Pi_0$ in post-processing (i.e. through the final term in \eqref{eq:logPORGR}). We then have
\begin{equation}\label{eq:postratio}
B^{m'}_0=\ln{\left(\frac{\Evid_{m'}}{\Evid_0}\right)}=\ln{\left[\frac{\Prob(m=m'|x,\Model)}{\Prob(m=0|x,\Model)}\right]},
\end{equation}
such that the Bayes factors in \eqref{eq:logPORGR} may be obtained directly from the sampled posterior distribution for $m$. A nested-sampling implementation of the product-space method was found to be viable by \cite{Hee2015} through application to a cosmological problem. In this paper, we use a similar implementation to compare a larger number of models in our GW test of GR, and perform an improved investigation of the errors on the obtained Bayes factors.

\subsection{Error estimation from rethreading}
\label{sec:rethreading}

If the Bayes factors in \eqref{eq:logPORGR} are obtained by evaluating the individual submodel evidences with regular nested sampling, the standard deviation of $\PORGR$ scales as $1/\sqrt{\Nlive}$ (from propagation of the log-evidence error \eqref{eq:logeviderror}). In the product-space method, however, the Bayes factors are ratios of posterior probabilities $p_m:=\Prob(m|x,\Model)$; computing them is then a parameter estimation problem, with a different associated uncertainty that depends on the posterior errors $\sigma_{p_m}$. These errors arise from the prior-mass approximation \eqref{eq:NSpriormass}, as in the case of $\sigma_{\ln{\Evid}}$, but also from the dimensional reduction in the likelihood reparametrisation $\Like(\theta)\to\Like(X)$ \citep{higson2017}. Nevertheless, \cite{higson2017} observe that $\sigma_{p_m}$ are determined by the relative errors of the nested-sampling weights, and hence are typically $<\sigma_{\ln{\Evid}}$. This is the key factor that underpins the improved efficiency of product-space nested sampling.

An algorithm for estimating the errors of $p_m$ (or of any quantity derived from $p_m$) over a single nested-sampling run has been proposed by \cite{higson2017}. The technique is motivated by the fact that any number of runs $r$ with $N_r$ live points may be merged into a single run with $\Nlive=\sum_rN_r$ \citep{Skilling2006}, by combining and reordering all of their dead points (and adjusting $\Nlive$ in \eqref{eq:NSpriormass} accordingly). Conversely, it is also possible to unravel a single nested-sampling run into its constituent ``threads'', i.e. a set of $\Nlive$ independent runs $r$ with $N_r=1$. This is achieved by tracking the birth and death order of the sampling points; each thread is then formed from the sequence of replacements for an original live point.

A distribution of runs with the original number of live points may be obtained rapidly from the set of threads, by using bootstrap resampling on the set and recombining (or ``rethreading'') the sample threads. These new runs contain only points that are present in the original run, but yield statistical variance in the estimated weights and $p_m$. In this work, we use the rethreading technique to generate $10^3$ realisations of $\PORGR$ from a single nested-sampling run, and thus to evaluate its mean and standard deviation directly. The rethreading results are also validated against those from repeated runs for the toy model in Section \ref{sec:toyresults}.

\section{Results: Sinusoidal toy model}
\label{sec:toyresults}

To demonstrate that product-space nested sampling with rethreading can explore qualitative EMRI likelihoods with improved efficiency, we first apply it to synthetic data generated from the toy EMRI model defined in Section \ref{sec:toy}. Two data sets $x$ are considered: one from the ``GR'' submodel $\Model_{0000}$, and one from the deformed submodel $\Model_{0010}$ with a ``$\mathcal{B}_3$ bump''. The signal parameters for $x_{0000}$ and $x_{0010}$ are $(A,\Omega)=(1,1)$ and $(A,\Omega,\lg{\epsilon_3})=(1,1,-1.9)$ respectively. Both waveforms are generated with duration $T_\mathrm{obs}=10^4$ (such that they contain $\sim10^3$ cycles) and a dimensionless sampling rate of unity, then renormalised to SNR $\rho=10$.

\begin{figure}
\centering
\begin{subfigure}[b]{\columnwidth}
\subcaption{$\lg{\epsilon_3}$--$\Omega$ slice}
\includegraphics[width=\columnwidth]{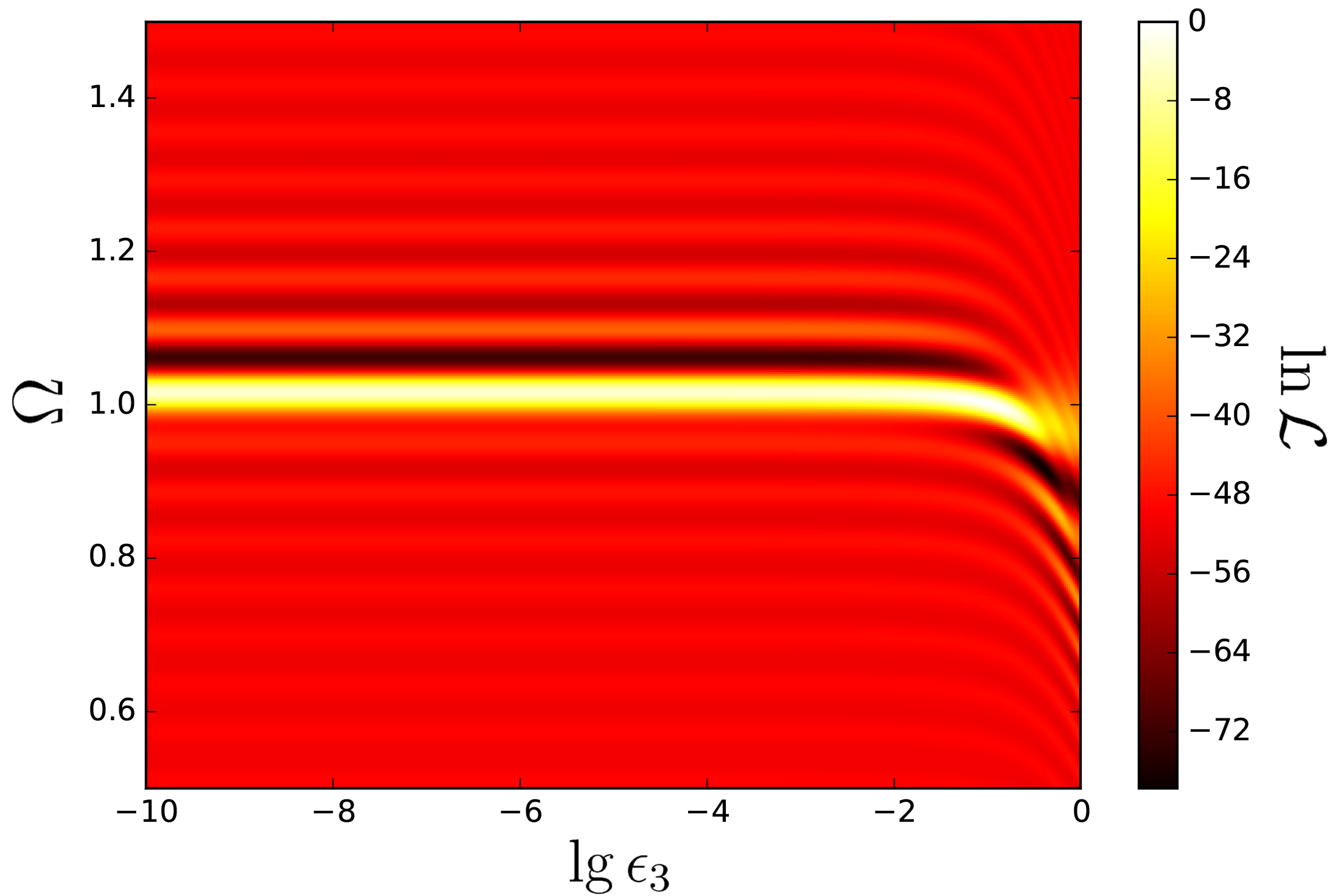}
\end{subfigure}
\par\bigskip
\begin{subfigure}[b]{\columnwidth}
\subcaption{$\lg{\epsilon_4}$--$\lg{\epsilon_5}$ slice}
\includegraphics[width=\columnwidth]{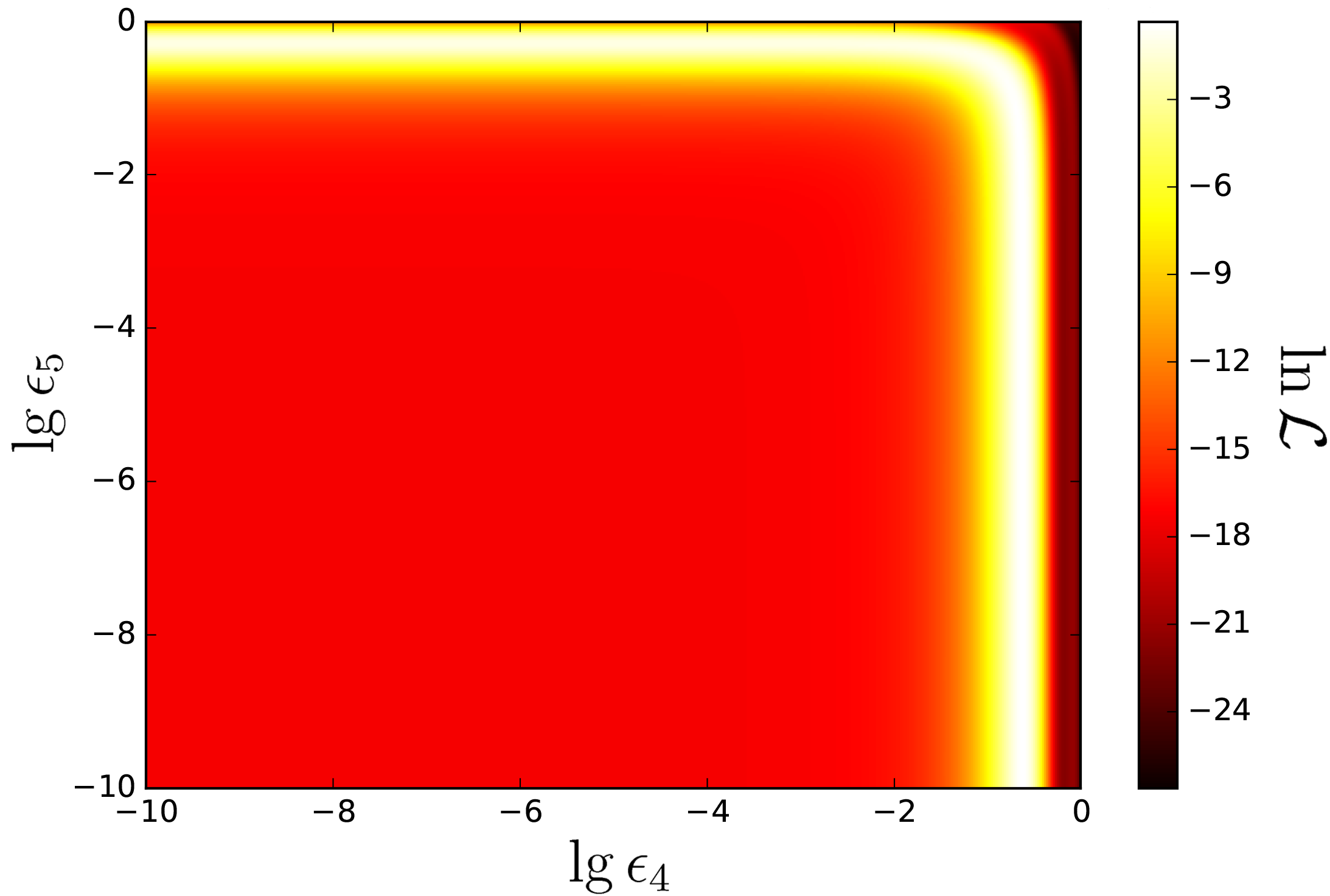}
\end{subfigure}
\caption{Log-likelihood slices for sinusoidal toy model: (a) submodel $\Model_{0010}$ with data $x_{0000}$; (b) submodel $\Model_{1100}$ with data $x_{0010}$.}
\label{fig:toylike}
\end{figure}

\begin{figure*}
\includegraphics[width=0.75\textwidth]{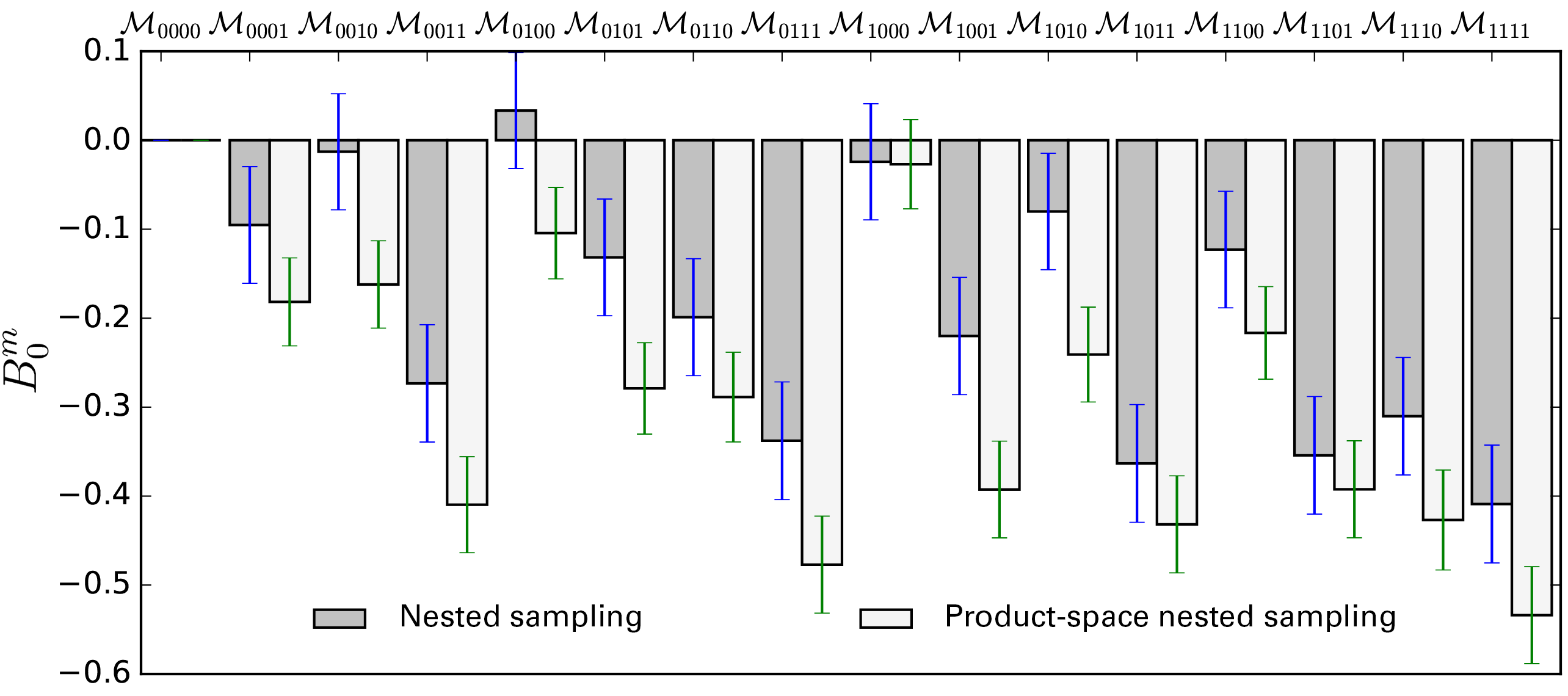}
\caption{Submodel Bayes factors $B^m_0$ ($B^0_0=0$ by definition) for regular and product-space nested sampling with sinusoidal toy data $x_{0000}$ (SNR $\rho=10$). Both methods have $\Nlive=2500$; regular nested sampling takes $4.2\times10^7$ likelihood calls in total, while product-space nested sampling takes $2.7\times10^6$ calls.}
\label{fig:toypost}
\end{figure*}

As the toy waveforms are near-sinusoidal, the GW likelihood \eqref{eq:like} contains damped oscillations in $\Omega$; to a lesser extent, such oscillations are also present for parameters that are highly correlated with frequency (e.g. $M$) in more realistic EMRI models. Figure \ref{fig:toylike}(a) shows the $\lg{\epsilon_3}$--$\Omega$ slice of $\ln{\Like}$ for the submodel $\Model_{0010}$ with the data $x_{0000}$, where $\Omega$ is seen to be relatively well localised in the moderate-SNR regime. However, the likelihood is highly degenerate in the deformation parameters. In the case of Figure \ref{fig:toylike}(a), any value for $\lg{\epsilon_3}$ that falls below some threshold $\approx-2$ ceases to deform the waveform significantly, and hence has negligible effect on the value of $\ln{\Like}$. This results in an extended region of high likelihood along the $\lg{\epsilon_3}$ axis, even though the true GR signal lies on the $\epsilon_3=0$ boundary of the $\Model_{0010}$ space.

The deformation parameters are also degenerate among themselves, which further complicates the likelihood surface if the true signal is deformed. For the data $x_{0010}$, the signal is not even contained in submodels without $\epsilon_3$; however, Figure \ref{fig:toylike}(b) shows that regions of high likelihood exist in the $\lg{\epsilon_4}$--$\lg{\epsilon_5}$ slice of $\ln{\Like}$ for $\Model_{1100}$, and that a $\lg{\epsilon_3}=-1.9$ deformation can be well approximated by a larger value of $\epsilon_4$ or $\epsilon_5$ (or a combination of the two). Although such degeneracies are partially broken by the additional degrees of freedom in more realistic generalised GW models, they are still expected to hamper parameter estimation for the individual deformations \citep{PhysRevD.85.082003,0264-9381-34-19-195009}. Nevertheless, they have a less severe impact on the null-hypothesis test in Section \ref{sec:test}, where the aim is not to distinguish among the deformed submodels.

For the two data sets, both regular and product-space nested sampling are used to obtain $\PORGR$ with an associated standard deviation $\sigma_P$. In the regular case, \eqref{eq:logPORGR} is evaluated piecewise by computing the individual submodel evidences, and the standard nested-sampling estimates for each log-evidence error \eqref{eq:logeviderror} are propagated to yield $\sigma_P$. For the product-space method, $\PORGR$ is obtained through a single nested-sampling run in the hypermodel space, and its approximate error is calculated in two ways: using the single-run rethreading technique, and from 50 repetitions of the same product-space run. The latter procedure is only to demonstrate that the methods in Sections \ref{sec:product} and \ref{sec:rethreading} are working as expected, and is not performed in Section \ref{sec:results}.

The product-space method manifestly provides computational savings over regular nested sampling in the null-hypothesis test, as it samples in just one parameter space instead of 16. However, the hypermodel space is more complex than a single submodel space and requires additional runtime to explore effectively, such that differences in efficiency cannot simply be estimated from the number of spaces sampled. To assess the gains when using the product-space method, we vary the \codeF{PolyChord} runtime parameter $\Nlive$ to obtain different degrees of precision on each corresponding evaluation of $\PORGR$, and compare the number of likelihood calls taken by the two methods to achieve the same $\sigma_P$. The other \codeF{PolyChord} runtime parameter is set as $\Nrep=30$, which has been chosen empirically to ensure the convergence of regular evidence estimates.

It is also instructive to study how regular and product-space nested sampling perform on the penultimate step in the evaluation of $\PORGR$, i.e. the individual submodel Bayes factors $B^m_0$. Figure \ref{fig:toypost} shows the $B^m_0$ and associated errors that are obtained from the two methods with $\Nlive=2500$, for the GR data $x_{0000}$. The Occam penalty on model complexity is clearly observed in both sets of results, as the relative evidence for each submodel decreases with the number of parameters it contains. However, the product-space method appears to systematically give Bayes factors that are more pronounced (negative), and in tension with the regular results. This is likely because the entire hypermodel space is explored with the same number of live points allocated to each submodel space in the regular method, leading to a slight degree of sampling bias. The Bayes factor errors for both methods are nevertheless comparable, since the errors on a posterior over $m$ are smaller than those on submodel evidence evaluations (as discussed in Section \ref{sec:rethreading}).

\begin{figure}
\centering
\begin{subfigure}[b]{\columnwidth}
\subcaption{Data $x_{0000}$}
\includegraphics[width=\columnwidth]{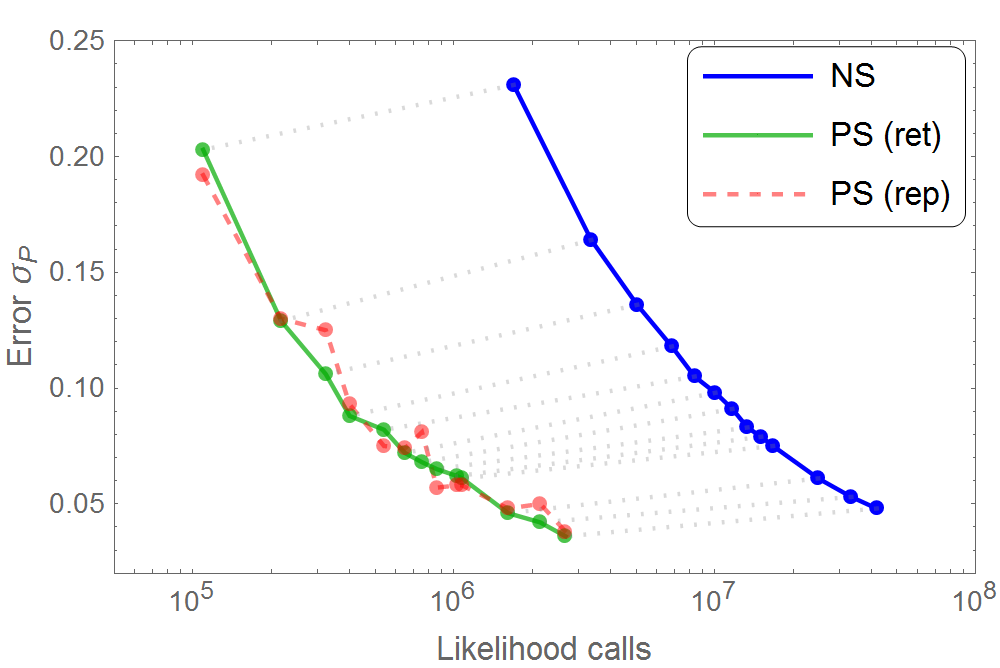}
\end{subfigure}
\par\bigskip
\begin{subfigure}[b]{\columnwidth}
\subcaption{Data $x_{0010}$}
\includegraphics[width=\columnwidth]{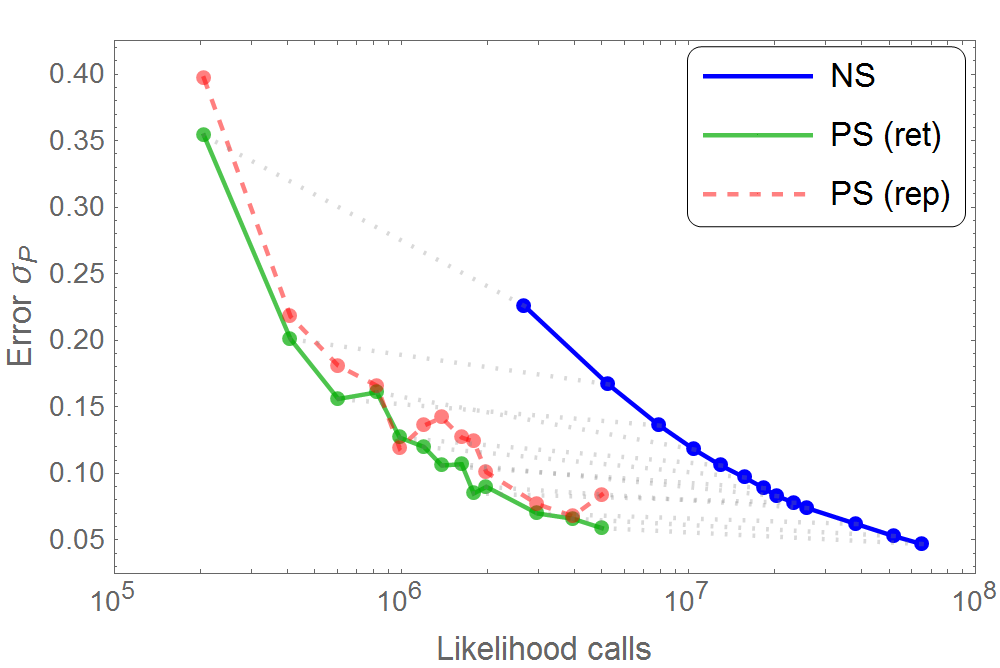}
\end{subfigure}
\caption{Error $\sigma_P$ of $\PORGR$ for (a) GR data $x_{0000}$ and (b) $\mathcal{B}_3$ data $x_{0010}$. Regular nested sampling (blue) is compared to product-space nested sampling with error estimates from single-run rethreading (green) and 50 repeated runs (red, dashed). Grey dotted lines indicate regular and product-space (rethreading) runs of equal $\Nlive$, ranging from 100 to 2500.}
\label{fig:toyprecision}
\end{figure}

\begin{figure}
\includegraphics[width=\columnwidth]{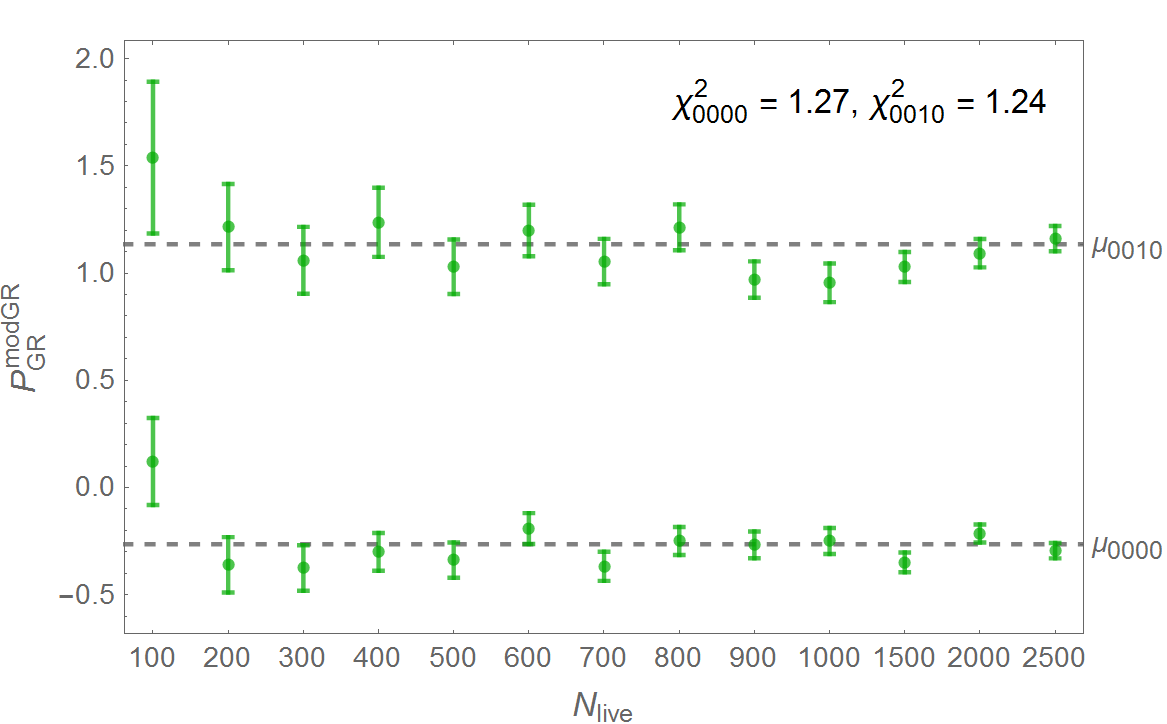}
\caption{Reduced chi-squared test of rethreading errors against sample mean $\mu$ of $\PORGR$, for GR data $x_{0000}$ (bottom) and $\mathcal{B}_3$ data $x_{0010}$ (top). Each sequence of 13 product-space runs with varying $\Nlive$ corresponds to a green curve in Figure \ref{fig:toyprecision}. The number of degrees of freedom is 12.}
\label{fig:toychi}
\end{figure}

In Figure \ref{fig:toyprecision}, the error $\sigma_P$ of $\PORGR$ for the two data sets $x_{0000}$ and $x_{0010}$ is plotted against the number of likelihood calls for a sequence of regular and product-space nested-sampling runs with $100\leq\Nlive\leq2500$. Both the rethreading and repetition error estimates for the product-space method are included; they are seen to agree well, with the latter showing more scatter (since they are computed from only 50 evaluations of $\PORGR$, as opposed to $10^3$ realisations in the rethreading technique). The rethreading error estimates are further validated through a reduced chi-squared test against the sample mean $\mu$ of $\PORGR$ in Figure \ref{fig:toychi}, where $\chi^2=\sum_i(P_i-\mu)^2/(12\sigma_P^2)\approx1$ for both data sets. We also find $\mu_{0000}\sim-1$ in the GR case and $\mu_{0010}\sim1$ in the $\mathcal{B}_3$ case, which is by design from our choices of SNR and $\epsilon_3$ for the synthetic data.

For both data sets, it is clear that product-space nested sampling is effective at reducing the computational cost required to reach a given level of precision (or alternatively, at providing greater precision with a given number of likelihood calls). In the GR case, the average gain in efficiency (i.e. the mean horizontal distance between the blue and green curves in Figure \ref{fig:toyprecision}) is a factor of around 24. Furthermore, as the likelihood surface over the hypermodel space is less complex for $x_{0000}$, nested sampling explores it nearly as efficiently as each of the 16 submodel spaces. This is seen by comparing regular and product-space runs of equal $\Nlive$, where the number of likelihood calls taken by the latter is almost exactly 16 times smaller, and its associated error is slightly lower as well.

In the $\mathcal{B}_3$ case, larger overall errors $\sigma_P$ are obtained for both methods, and the average reduction in computational cost with product-space nested sampling is reduced to a factor of around 9. As expected, this is largely caused by the increased complexity of the likelihood surface over the hypermodel space for $x_{0010}$: additional modes are present in the parameter spaces of the eight submodels containing $\epsilon_3$, as well as in other submodel spaces due to deformation parameter degeneracies such as that from Figure \ref{fig:toylike}(b). When comparing regular and product-space runs of equal $\Nlive$, the number of likelihood calls taken by the latter is only 13 times smaller, and its associated error is now slightly higher. It follows from these results that computational savings will likely be smallest for data generated from $\Model_{1111}$; we discuss this in Section \ref{sec:results}.

Another contribution to the overall difference in $\sigma_P$ between the two data sets arises from the construction of $\PORGR$ itself. For the $\mathcal{B}_3$ data, the parameter space of the null submodel $\Model_{0000}$ is a region of lower posterior probability in the product-space approach, and hence the relative posterior error on $\Prob(m=0)$ is higher. This error propagates into every submodel Bayes factor $B^m_0$ via \eqref{eq:postratio}, which increases the final error $\sigma_P$. The effect is also present for regular nested sampling, since $\Model_{0000}$ has a higher relative evidence error as well. However, the increase in $\sigma_P$ only becomes significant in the strong-deformation regime, which is unlikely for actual tests of GR; furthermore, it can also be mitigated in practice (e.g. by using the actual submodel index prior $\sum_{m\neq0}\Pi_m=\Pi_0$ when sampling).

\section{Results: Bumpy analytic kludge model}
\label{sec:results}

For a more realistic generalised EMRI waveform model in the product-space approach, the dimensionality of the hypermodel parameter space is considerably higher. In the case of the bAK model, $\Theta=\Theta_\mathrm{AK}\cup\Theta_\mathcal{B}\cup\{m\}$ is 19-dimensional and parametrised by 14 GR parameters, four deformation parameters, and the submodel index. The likelihood surface over the GR parameter space is known to be highly multimodal with a large information content \citep{0264-9381-21-20-003}. A full exploration of this space is hampered by the significant computational cost of waveform generation (even though the AK formalism already provides the cheapest EMRI waveforms available), and is beyond the scope of the present work. We instead fix all but seven of the hypermodel parameters, allowing only the component masses, deformation parameters and submodel index to vary. For our synthetic data, the intrinsic GR parameters \eqref{eq:intrinsic} of the signal are chosen as $\Theta_\mathrm{int}=(1,6,0.1,0.1,0.9,0,0)$; the waveform is two months long with an initial frequency of $2\,\mathrm{mHz}$ (such that it contains $\sim10^4$ cycles), and is sampled at $0.2\,\mathrm{Hz}$.

\begin{figure*}
\includegraphics[width=0.75\textwidth]{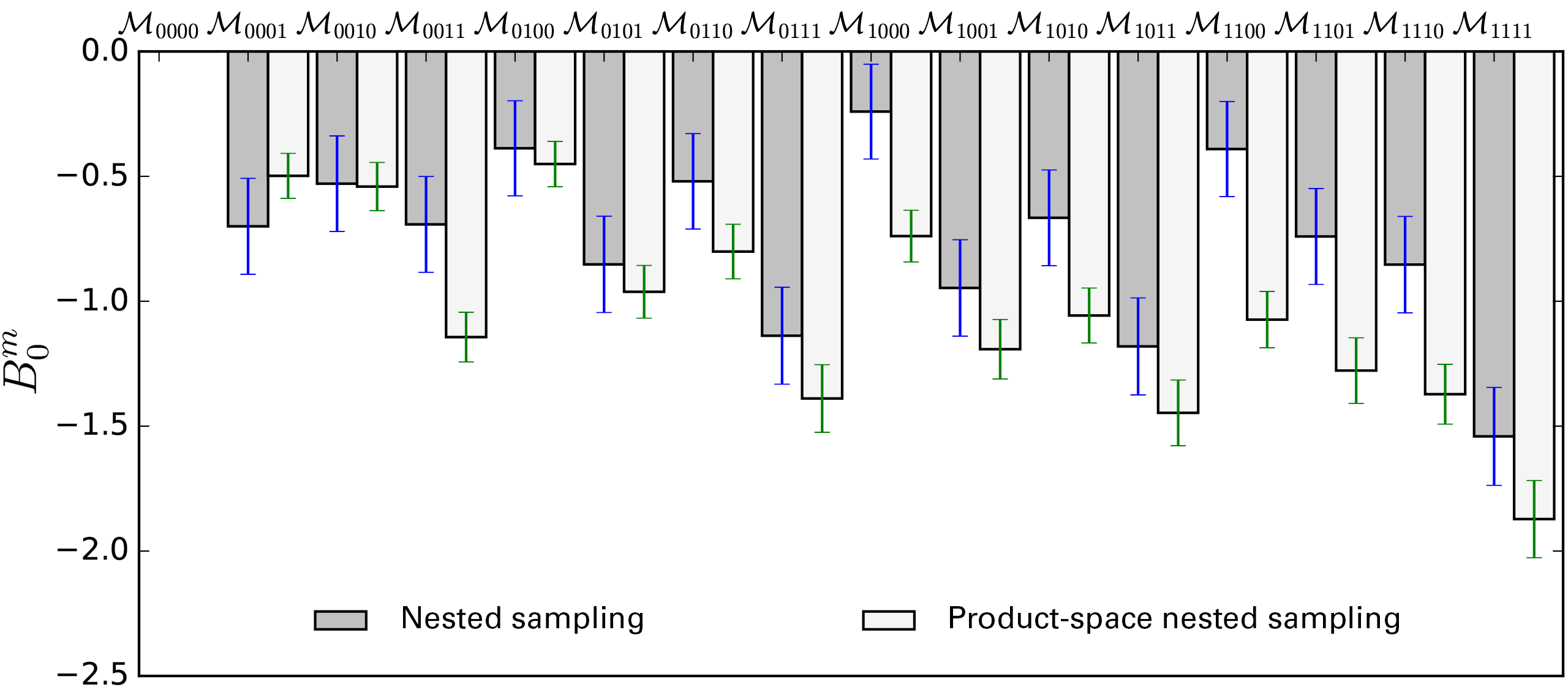}
\caption{Submodel Bayes factors $B^m_0$ ($B^0_0=0$ by definition) for regular and product-space nested sampling with bAK data $x_{0000}$ (SNR $\rho=10$). Both methods have $\Nlive=500$; regular nested sampling takes $1.2\times10^6$ likelihood calls in total, while product-space nested sampling takes $7.8\times10^5$ calls.}
\label{fig:post}
\end{figure*}

Instead of considering synthetic data from a deformed submodel as in Section \ref{sec:toyresults}, we restrict analysis here to the more realistic case of a GR signal, and investigate the effect of SNR on sampling performance. The two data sets studied in this section are both generated from $\Model_{0000}$ with the same GR parameters, but are renormalised to SNRs $\rho=10$ and $\rho=100$ respectively. Figure \ref{fig:post} shows the $B^m_0$ and associated errors that are obtained from regular and product-space nested sampling with $\Nlive=500$, for the moderate-SNR case $\rho=10$. The Occam penalty is again apparent, but the submodel Bayes factors are generally lower than in Figure \ref{fig:toypost} even though the SNR is the same; this is likely due to the EMRI's orbital evolution reducing degeneracy in the deformed parameters. As in Figure \ref{fig:toypost}, there also appears to be a slight systematic difference between the Bayes factors from the two methods. Both the regular and product-space methods correctly favour the null submodel $\Model_{0000}$, although the latter does so with smaller errors and fewer likelihood calls for the same number of live points.

\begin{figure}
\includegraphics[width=\columnwidth]{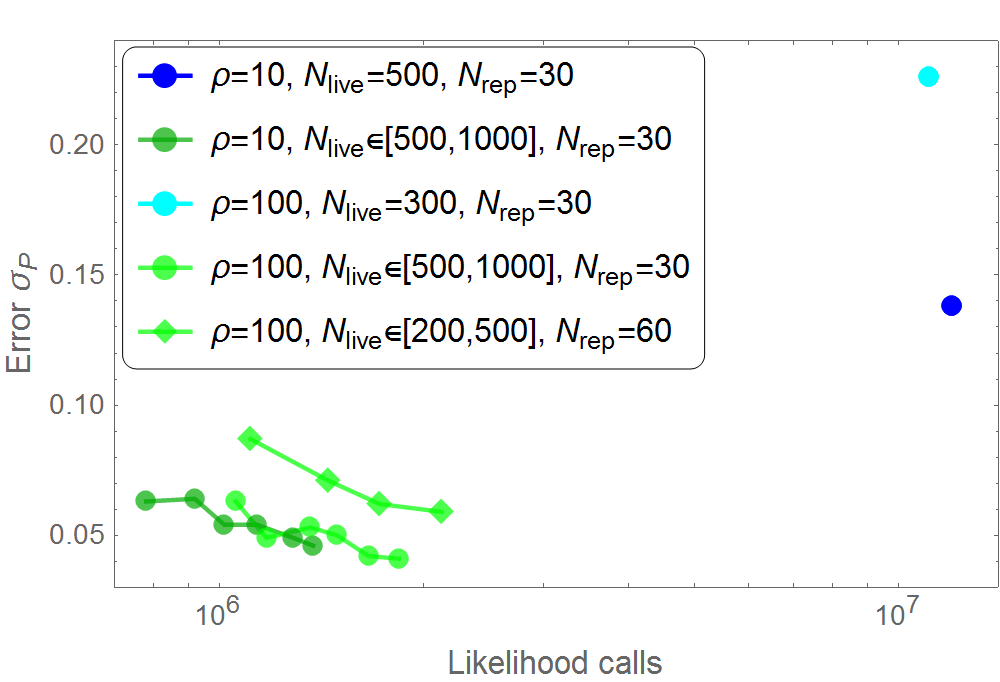}
\caption{Error $\sigma_P$ of $\PORGR$ for GR data $x_{0000}$ with varying SNR $\rho\in\{10,100\}$, sampling resolution $\Nlive\in[200,1000]$, and sampling reliability $\Nrep\in\{30,60\}$. Regular nested sampling (blue) is compared to product-space nested sampling with rethreading (green).}
\label{fig:precision}
\end{figure}

In Figure \ref{fig:precision}, the error $\sigma_P$ of $\PORGR$ for the $\rho=10$ and $\rho=100$ data sets is plotted against the number of likelihood calls for several product-space nested-sampling runs with varying \codeF{PolyChord} runtime parameters $\Nlive$ and $\Nrep$. These are compared against a single regular nested-sampling run for each SNR value (due to the considerably higher computational cost of the regular method). It is clear that the efficiency gains obtained for the sinusoidal toy model are still present for the bAK model, in that every product-space run shown has both better precision and lower computational cost than the two regular runs. As indicated by the results in Section \ref{sec:toyresults}, these gains might be diminished in the case of data that is generated from more complex submodels such as $\Model_{1111}$. However, since there is strong prior expectation for an actual EMRI signal to be well described by GR, the order-of-magnitude savings observed here will likely be close to what is obtained in practice.

Furthermore, since the log-evidence error \eqref{eq:logeviderror} scales as $1/\sqrt{\Nlive}$ and the number of likelihood calls increases approximately linearly with $\Nlive$, an extrapolation of both regular nested-sampling runs to 5\% error ($\sim10^8$ calls) indicates that for the bAK likelihood, the reduction of computational cost from the product-space method is boosted to around two orders of magnitude. A direct verification of this statement is impractical for the present work, since the cost of each $\sim10^7$-call run in Figure \ref{fig:precision} is $\approx4000$ core hours (a single call to the bAK likelihood takes $\approx1.5\,\mathrm{s}$). Nevertheless, the scaling of errors on nested-sampling evidences is both well understood and reliable (as indicated by the smoothness of the blue curves in Figure \ref{fig:toyprecision}), which lends credence to the validity of such an extrapolation.

\begin{figure}
\includegraphics[width=\columnwidth]{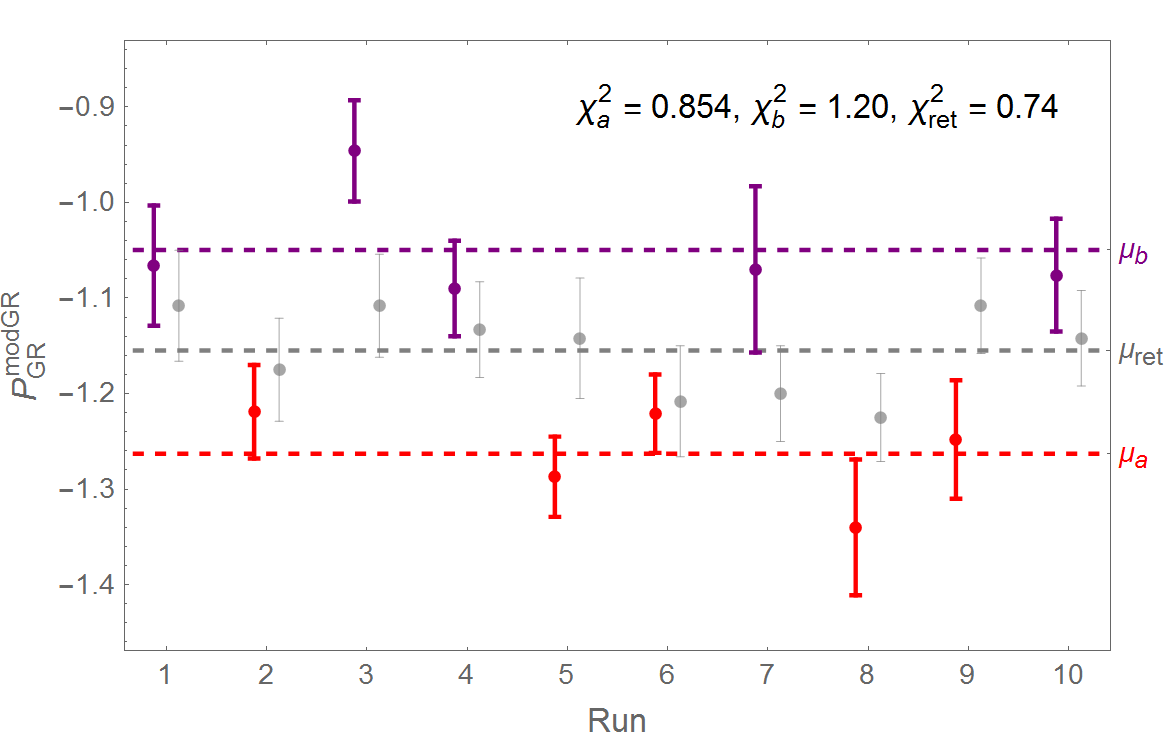}
\caption{Reduced chi-squared test of product-space errors, for GR data $x_{0000}$ with SNR $\rho=100$. As the 10 original runs have inadequate sampling accuracy, they appear to be clustered around two distinct values $\mu_a$ (red) and $\mu_b$ (purple). Unravelling and rethreading them into 10 runs of equal $\Nlive$ reduces the bias, and yields a single value $\mu_\mathrm{ret}$ (grey) as expected.}
\label{fig:chi}
\end{figure}

Significant sampling bias is introduced into the results for the high-SNR case $\rho=100$; this is because optima in the likelihood surface become more localised, and hence require higher sampling resolution/reliability to map out accurately. The values of $\PORGR$ and their associated errors from the 10 high-SNR product-space runs (indicated by the light green points in Figure \ref{fig:precision}) are plotted in Figure \ref{fig:chi}. Computing the reduced chi-squared statistic for these runs gives $\chi^2\approx5$, such that $\sigma_P$ for any given run appears to underrepresent the observed scatter within the set of runs. Since the rethreading technique has been validated in Section \ref{sec:toyresults}, the large $\chi^2$ suggests that the hypermodel space is not being sampled consistently between runs. If any particular run misses a region of high posterior probability, then its threads will not contain enough information to provide a rethreading estimate that represents its true error; this could occur if $\Nlive$ is not large enough to find sharply defined modes, or if replacement live points are correlated with discarded ones due to inadequate $\Nrep$. The sampling bias observed here (and to a lesser extent in Figures \ref{fig:toypost} and \ref{fig:post}) might be better characterised with new diagnostic tests \citep{higson2018} in future work.

In Figure \ref{fig:chi}, we illustrate through a clustered chi-squared test that the obtained $\PORGR$ and $\sigma_P$ actually show a closer fit to two distinct sampling distributions with differing mean (since $\chi^2\approx1$ for each cluster). This analysis is not to be taken at face value, since $\PORGR$ clearly has a unique underlying value. It does however highlight the possibility of drawing an erroneous conclusion from the null-hypothesis test due to sampling bias, especially if the runtime parameters $\Nlive$ and $\Nrep$ are set too low to adequately explore high-SNR likelihoods. Furthermore, with the added complexity of the EMRI likelihood surface, it is computationally unfeasible to systematically increase these parameters without bound. We thus propose a strategy for making the product-space method robust to sampling bias; the idea is simple, and made possible by the same premise that facilitates the rethreading technique.

As discussed in Section \ref{sec:rethreading}, nested sampling permits both the unravelling and interweaving of independent runs. Although threads from a single run might have correlations that reflect the sampling bias present in that run, such correlations are more diffuse in a set of threads from a large population of runs, and sampling bias will be reduced in runs that are randomly reconstructed from these threads. Hence the best way to utilise all of the information in the 10 high-SNR product-space runs is to unravel them into their constituent threads (5900 of them in this case) and rethread these into realisations of a run with $\Nlive=5900$, which produces the value $\PORGR=-1.16\pm0.02$. The error is far smaller than that of any run in Figure \ref{fig:precision}, which is perhaps unsurprising given the combination of information from all the runs. To demonstrate that this procedure does actually serve to decorrelate the individual threads, 10 new runs with $\Nlive=590$ are constructed through rethreading; their results (the grey points in Figure \ref{fig:chi}) are seen to exhibit none of the sampling bias present in the original runs.

\section{Conclusion}
\label{sec:conclusion}

In this paper, we have adapted, combined and assessed a variety of recent modelling/statistical techniques to devise a preliminary framework for testing GR with EMRI observations from future space-based GW detectors. A generalised EMRI waveform model \citep{PhysRevD.84.064016} and its toy surrogate are trialled in a null-hypothesis test developed for LIGO sources \citep{PhysRevD.85.082003}; the method of product-space nested sampling \citep{Hee2015} with rethreading error estimates \citep{higson2017} is shown to systematically increase computational efficiency by an order of magnitude over regular evidence-based sampling.

The results and observations presented here are quite general; they are relevant not just for the outlined EMRI test of GR, but indeed any similar parametrised test that uses a generalised waveform model to describe GW sources in modified gravity (although the need to reduce computational cost is most strongly motivated for EMRIs). Product-space nested sampling with rethreading is furthermore shown to be efficient, robust and hence potentially useful for a broad range of model selection problems beyond the null-hypothesis test in this work. While much of the present analysis is exclusive to nested-sampling theory, some results (e.g. the characterisation of posterior versus evidence errors) might also be applicable to other algorithms such as product-space MCMC.

Although the computational savings afforded by our proposed methods are promising, this work is only the first step in developing a practical infrastructure for testing GR with future EMRI observations. The framework should eventually incorporate other techniques for increased efficiency, such as reduced-order quadratures \citep{PhysRevD.87.124005} to accelerate individual likelihood evaluations, or dynamic nested sampling \citep{2017arXiv170403459H} to improve sampling convergence. Finally, the actual accuracy and instructiveness of the null-hypothesis test must also be validated on data sets containing realistic source signals and detector noise, e.g. as performed by \cite{PhysRevD.97.044033} for the constraints on deformation parameters in the LIGO test, but with additional focus on the final posterior odds ratio.

\section*{Acknowledgements}

We thank Nicolas Yunes, Michele Vallisneri and Stephen Taylor for useful comments on the manuscript. AJKC acknowledges support from the Jet Propulsion Laboratory (JPL) Research and Technology Development programme. SH thanks the Science and Technology Facilities Council (STFC) for financial support. CJM acknowledges financial support provided under the European Union's H2020 ERC Consolidator Grant ``Matter and strong-field gravity: New frontiers in Einstein's theory'' grant agreement no. MaGRaTh646597, and networking support by the COST Action CA16104. Parts of this work were performed using the Darwin Supercomputer of the University of Cambridge High Performance Computing Service (\href{http://www.hpc.cam.ac.uk/}{http://www.hpc.cam.ac.uk/}), provided by Dell Inc. using Strategic Research Infrastructure Funding from the Higher Education Funding Council for England and funding from STFC. Parts of this work were also undertaken on the COSMOS Shared Memory system at DAMTP, University of Cambridge operated on behalf of the STFC DiRAC HPC Facility; this equipment is funded by BIS National E-infrastructure capital grant ST/J005673/1 and STFC grants ST/H008586/1, ST/K00333X/1. Parts of this work were also carried out at JPL, California Institute of Technology, under a contract with the National Aeronautics and Space Administration. \textcopyright{2018}. All rights reserved.

\bibliographystyle{mnras}
\bibliography{references}

\begin{thebibliography}{}
\makeatletter
\relax
\def\mn@urlcharsother{\let\do\@makeother \do\$\do\&\do\#\do\^\do\_\do\%\do\~}
\def\mn@doi{\begingroup\mn@urlcharsother \@ifnextchar [ {\mn@doi@}
  {\mn@doi@[]}}
\def\mn@doi@[#1]#2{\def\@tempa{#1}\ifx\@tempa\@empty \href
  {http://dx.doi.org/#2} {doi:#2}\else \href {http://dx.doi.org/#2} {#1}\fi
  \endgroup}
\def\mn@eprint#1#2{\mn@eprint@#1:#2::\@nil}
\def\mn@eprint@arXiv#1{\href {http://arxiv.org/abs/#1} {{\tt arXiv:#1}}}
\def\mn@eprint@dblp#1{\href {http://dblp.uni-trier.de/rec/bibtex/#1.xml}
  {dblp:#1}}
\def\mn@eprint@#1:#2:#3:#4\@nil{\def\@tempa {#1}\def\@tempb {#2}\def\@tempc
  {#3}\ifx \@tempc \@empty \let \@tempc \@tempb \let \@tempb \@tempa \fi \ifx
  \@tempb \@empty \def\@tempb {arXiv}\fi \@ifundefined
  {mn@eprint@\@tempb}{\@tempb:\@tempc}{\expandafter \expandafter \csname
  mn@eprint@\@tempb\endcsname \expandafter{\@tempc}}}

\bibitem[\protect\citeauthoryear{Abbott et~al.}{Abbott
  et~al.}{2016a}]{PhysRevX.6.041015}
Abbott B.~P.,  et~al., 2016a, \mn@doi [Phys. Rev. X]
  {10.1103/PhysRevX.6.041015}, 6, 041015

\bibitem[\protect\citeauthoryear{Abbott et~al.}{Abbott
  et~al.}{2016b}]{PhysRevLett.116.221101}
Abbott B.~P.,  et~al., 2016b, \mn@doi [Phys. Rev. Lett.]
  {10.1103/PhysRevLett.116.221101}, 116, 221101

\bibitem[\protect\citeauthoryear{Abbott et~al.}{Abbott
  et~al.}{2017a}]{PhysRevLett.118.221101}
Abbott B.~P.,  et~al., 2017a, \mn@doi [Phys. Rev. Lett.]
  {10.1103/PhysRevLett.118.221101}, 118, 221101

\bibitem[\protect\citeauthoryear{Abbott et~al.}{Abbott
  et~al.}{2017b}]{PhysRevLett.119.141101}
Abbott B.~P.,  et~al., 2017b, \mn@doi [Phys. Rev. Lett.]
  {10.1103/PhysRevLett.119.141101}, 119, 141101

\bibitem[\protect\citeauthoryear{Abbott et~al.}{Abbott
  et~al.}{2017c}]{PhysRevLett.119.161101}
Abbott B.~P.,  et~al., 2017c, \mn@doi [Phys. Rev. Lett.]
  {10.1103/PhysRevLett.119.161101}, 119, 161101

\bibitem[\protect\citeauthoryear{Abbott et~al.}{Abbott
  et~al.}{2017d}]{2017ApJ...851L..35A}
Abbott B.~P.,  et~al., 2017d, \mn@doi [\apjl] {10.3847/2041-8213/aa9f0c}, \href
  {http://adsabs.harvard.edu/abs/2017ApJ...851L..35A} {851, L35}

\bibitem[\protect\citeauthoryear{Abbott et~al.}{Abbott
  et~al.}{2018a}]{2018arXiv180210194T}
Abbott B.~P.,  et~al., 2018a, preprint, \href
  {http://adsabs.harvard.edu/abs/2018arXiv180210194T} {} (\mn@eprint {arXiv}
  {1802.10194})

\bibitem[\protect\citeauthoryear{Abbott et~al.}{Abbott
  et~al.}{2018b}]{PhysRevLett.120.031104}
Abbott B.~P.,  et~al., 2018b, \mn@doi [Phys. Rev. Lett.]
  {10.1103/PhysRevLett.120.031104}, 120, 031104

\bibitem[\protect\citeauthoryear{Agathos, Del~Pozzo, Li, Van Den~Broeck, Veitch
   \& Vitale}{Agathos et~al.}{2014}]{PhysRevD.89.082001}
Agathos M.,  Del~Pozzo W.,  Li T. G.~F.,  Van Den~Broeck C.,  Veitch J.,
  Vitale S.,  2014, \mn@doi [Phys. Rev. D] {10.1103/PhysRevD.89.082001}, 89,
  082001

\bibitem[\protect\citeauthoryear{Amaro-Seoane et~al.,}{Amaro-Seoane
  et~al.}{2012}]{0264-9381-29-12-124016}
Amaro-Seoane P.,  et~al., 2012, Classical and Quantum Gravity, 29, 124016

\bibitem[\protect\citeauthoryear{{Amaro-Seoane} et~al.,}{{Amaro-Seoane}
  et~al.}{2013}]{2013GWN.....6....4A}
{Amaro-Seoane} P.,  et~al., 2013, GW Notes, Vol.~6, p.~4-110, \href
  {http://adsabs.harvard.edu/abs/2013GWN.....6....4A} {6, 4}

\bibitem[\protect\citeauthoryear{{Amaro-Seoane} et~al.,}{{Amaro-Seoane}
  et~al.}{2017}]{2017arXiv170200786A}
{Amaro-Seoane} P.,  et~al., 2017, preprint, \href
  {http://adsabs.harvard.edu/abs/2017arXiv170200786A} {} (\mn@eprint {arXiv}
  {1702.00786})

\bibitem[\protect\citeauthoryear{Apostolatos, Cutler, Sussman  \&
  Thorne}{Apostolatos et~al.}{1994}]{PhysRevD.49.6274}
Apostolatos T.~A.,  Cutler C.,  Sussman G.~J.,   Thorne K.~S.,  1994, \mn@doi
  [Phys. Rev. D] {10.1103/PhysRevD.49.6274}, 49, 6274

\bibitem[\protect\citeauthoryear{Babak et~al.,}{Babak
  et~al.}{2017}]{PhysRevD.95.103012}
Babak S.,  et~al., 2017, \mn@doi [Phys. Rev. D] {10.1103/PhysRevD.95.103012},
  95, 103012

\bibitem[\protect\citeauthoryear{Barack \& Cutler}{Barack \&
  Cutler}{2004}]{PhysRevD.69.082005}
Barack L.,  Cutler C.,  2004, \mn@doi [Phys. Rev. D]
  {10.1103/PhysRevD.69.082005}, 69, 082005

\bibitem[\protect\citeauthoryear{Barack \& Cutler}{Barack \&
  Cutler}{2007}]{PhysRevD.75.042003}
Barack L.,  Cutler C.,  2007, \mn@doi [Phys. Rev. D]
  {10.1103/PhysRevD.75.042003}, 75, 042003

\bibitem[\protect\citeauthoryear{Barker \& O'Connell}{Barker \&
  O'Connell}{1975}]{PhysRevD.12.329}
Barker B.~M.,  O'Connell R.~F.,  1975, \mn@doi [Phys. Rev. D]
  {10.1103/PhysRevD.12.329}, 12, 329

\bibitem[\protect\citeauthoryear{Benenti \& Francaviglia}{Benenti \&
  Francaviglia}{1979}]{Benenti1979}
Benenti S.,  Francaviglia M.,  1979, \mn@doi [General Relativity and
  Gravitation] {10.1007/BF00757025}, 10, 79

\bibitem[\protect\citeauthoryear{Canizares, Field, Gair  \& Tiglio}{Canizares
  et~al.}{2013}]{PhysRevD.87.124005}
Canizares P.,  Field S.~E.,  Gair J.~R.,   Tiglio M.,  2013, \mn@doi [Phys.
  Rev. D] {10.1103/PhysRevD.87.124005}, 87, 124005

\bibitem[\protect\citeauthoryear{Carlin \& Chib}{Carlin \&
  Chib}{1995}]{10.2307/2346151}
Carlin B.~P.,  Chib S.,  1995, Journal of the Royal Statistical Society. Series
  B (Methodological), 57, 473

\bibitem[\protect\citeauthoryear{Chib}{Chib}{1995}]{10.2307/2291521}
Chib S.,  1995, Journal of the American Statistical Association, 90, 1313

\bibitem[\protect\citeauthoryear{Chua \& Gair}{Chua \&
  Gair}{2015}]{0264-9381-32-23-232002}
Chua A. J.~K.,  Gair J.~R.,  2015, \mn@doi [Classical and Quantum Gravity]
  {10.1088/0264-9381/32/23/232002}, 32, 232002

\bibitem[\protect\citeauthoryear{Chua, Moore  \& Gair}{Chua
  et~al.}{2017}]{PhysRevD.96.044005}
Chua A. J.~K.,  Moore C.~J.,   Gair J.~R.,  2017, \mn@doi [Phys. Rev. D]
  {10.1103/PhysRevD.96.044005}, 96, 044005

\bibitem[\protect\citeauthoryear{Collins \& Hughes}{Collins \&
  Hughes}{2004}]{PhysRevD.69.124022}
Collins N.~A.,  Hughes S.~A.,  2004, \mn@doi [Phys. Rev. D]
  {10.1103/PhysRevD.69.124022}, 69, 124022

\bibitem[\protect\citeauthoryear{Consonni, Forster  \& La~Rocca}{Consonni
  et~al.}{2013}]{consonni2013}
Consonni G.,  Forster J.~J.,   La~Rocca L.,  2013, \mn@doi [Statist. Sci.]
  {10.1214/13-STS433}, 28, 398

\bibitem[\protect\citeauthoryear{Cutler}{Cutler}{1998}]{PhysRevD.57.7089}
Cutler C.,  1998, \mn@doi [Phys. Rev. D] {10.1103/PhysRevD.57.7089}, 57, 7089

\bibitem[\protect\citeauthoryear{Cutler \& Flanagan}{Cutler \&
  Flanagan}{1994}]{PhysRevD.49.2658}
Cutler C.,  Flanagan E.~E.,  1994, \mn@doi [Phys. Rev. D]
  {10.1103/PhysRevD.49.2658}, 49, 2658

\bibitem[\protect\citeauthoryear{Dooley, Akutsu, Dwyer  \& Puppo}{Dooley
  et~al.}{2015}]{1742-6596-610-1-012012}
Dooley K.~L.,  Akutsu T.,  Dwyer S.,   Puppo P.,  2015, Journal of Physics:
  Conference Series, 610, 012012

\bibitem[\protect\citeauthoryear{Feroz, Gair, Hobson  \& Porter}{Feroz
  et~al.}{2009a}]{0264-9381-26-21-215003}
Feroz F.,  Gair J.~R.,  Hobson M.~P.,   Porter E.~K.,  2009a, Classical and
  Quantum Gravity, 26, 215003

\bibitem[\protect\citeauthoryear{Feroz, Hobson  \& Bridges}{Feroz
  et~al.}{2009b}]{Feroz2009}
Feroz F.,  Hobson M.~P.,   Bridges M.,  2009b, \mnras, 398, 1601

\bibitem[\protect\citeauthoryear{{Feroz}, {Hobson}, {Cameron}  \&
  {Pettitt}}{{Feroz} et~al.}{2013}]{Feroz2013}
{Feroz} F.,  {Hobson} M.~P.,  {Cameron} E.,   {Pettitt} A.~N.,  2013, preprint
  (\mn@eprint {arXiv} {1306.2144})

\bibitem[\protect\citeauthoryear{Gair \& Yunes}{Gair \&
  Yunes}{2011}]{PhysRevD.84.064016}
Gair J.,  Yunes N.,  2011, \mn@doi [Phys. Rev. D] {10.1103/PhysRevD.84.064016},
  84, 064016

\bibitem[\protect\citeauthoryear{Gair, Barack, Creighton, Cutler, Larson,
  Phinney  \& Vallisneri}{Gair et~al.}{2004}]{0264-9381-21-20-003}
Gair J.~R.,  Barack L.,  Creighton T.,  Cutler C.,  Larson S.~L.,  Phinney
  E.~S.,   Vallisneri M.,  2004, Classical and Quantum Gravity, 21, S1595

\bibitem[\protect\citeauthoryear{Gair, Vallisneri, Larson  \& Baker}{Gair
  et~al.}{2013}]{Gair2013}
Gair J.~R.,  Vallisneri M.,  Larson S.~L.,   Baker J.~G.,  2013, \mn@doi
  [Living Reviews in Relativity] {10.12942/lrr-2013-7}, 16, 7

\bibitem[\protect\citeauthoryear{Gelman \& Meng}{Gelman \&
  Meng}{1998}]{Gelman1998}
Gelman A.,  Meng X.-L.,  1998, Stat. Sci., 13, 163

\bibitem[\protect\citeauthoryear{George \& Foster}{George \&
  Foster}{2000}]{doi:10.1093/biomet/87.4.731}
George E.,  Foster D.~P.,  2000, \mn@doi [Biometrika]
  {10.1093/biomet/87.4.731}, 87, 731

\bibitem[\protect\citeauthoryear{Glampedakis \& Babak}{Glampedakis \&
  Babak}{2006}]{0264-9381-23-12-013}
Glampedakis K.,  Babak S.,  2006, Classical and Quantum Gravity, 23, 4167

\bibitem[\protect\citeauthoryear{Godsill}{Godsill}{2001}]{doi:10.1198/10618600152627924}
Godsill S.~J.,  2001, \mn@doi [Journal of Computational and Graphical
  Statistics] {10.1198/10618600152627924}, 10, 230

\bibitem[\protect\citeauthoryear{Green}{Green}{1995}]{doi:10.1093/biomet/82.4.711}
Green P.~J.,  1995, \mn@doi [Biometrika] {10.1093/biomet/82.4.711}, 82, 711

\bibitem[\protect\citeauthoryear{{Handley}, {Hobson}  \& {Lasenby}}{{Handley}
  et~al.}{2015a}]{Handley2015}
{Handley} W.~J.,  {Hobson} M.~P.,   {Lasenby} A.~N.,  2015a, \mnras, 450, L61

\bibitem[\protect\citeauthoryear{{Handley}, {Hobson}  \& {Lasenby}}{{Handley}
  et~al.}{2015b}]{Handley2015b}
{Handley} W.~J.,  {Hobson} M.~P.,   {Lasenby} A.~N.,  2015b, \mnras, 453, 4384

\bibitem[\protect\citeauthoryear{Hee, Handley, Hobson  \& Lasenby}{Hee
  et~al.}{2015}]{Hee2015}
Hee S.,  Handley W.~J.,  Hobson M.~P.,   Lasenby A.~N.,  2015, \mnras, 455,
  2461

\bibitem[\protect\citeauthoryear{Higson, Handley, Hobson  \& Lasenby}{Higson
  et~al.}{2017}]{higson2017}
Higson E.,  Handley W.,  Hobson M.,   Lasenby A.,  2017, \mn@doi [Bayesian
  Analysis] {10.1214/17-BA1075}

\bibitem[\protect\citeauthoryear{{Higson}, {Handley}, {Hobson}  \&
  {Lasenby}}{{Higson} et~al.}{2018a}]{higson2018}
{Higson} E.,  {Handley} W.,  {Hobson} M.,   {Lasenby} A.,  2018a, preprint
  (\mn@eprint {arXiv} {1804.06406})

\bibitem[\protect\citeauthoryear{{Higson}, {Handley}, {Hobson}  \&
  {Lasenby}}{{Higson} et~al.}{2018b}]{2017arXiv170403459H}
{Higson} E.,  {Handley} W.,  {Hobson} M.,   {Lasenby} A.,  2018b, preprint,
  \href {http://adsabs.harvard.edu/abs/2017arXiv170403459H} {} (\mn@eprint
  {arXiv} {1704.03459})

\bibitem[\protect\citeauthoryear{Hughes}{Hughes}{2000}]{PhysRevD.61.084004}
Hughes S.~A.,  2000, \mn@doi [Phys. Rev. D] {10.1103/PhysRevD.61.084004}, 61,
  084004

\bibitem[\protect\citeauthoryear{Jeffreys}{Jeffreys}{1961}]{Jeffreys1961}
Jeffreys H.,  1961, {The Theory of Probability}.
Oxford University Press

\bibitem[\protect\citeauthoryear{{Junker} \& {Schaefer}}{{Junker} \&
  {Schaefer}}{1992}]{1992MNRAS.254..146J}
{Junker} W.,  {Schaefer} G.,  1992, \mn@doi [\mnras] {10.1093/mnras/254.1.146},
  \href {http://adsabs.harvard.edu/abs/1992MNRAS.254..146J} {254, 146}

\bibitem[\protect\citeauthoryear{Kass \& Raftery}{Kass \&
  Raftery}{1995}]{doi:10.1080/01621459.1995.10476572}
Kass R.~E.,  Raftery A.~E.,  1995, \mn@doi [Journal of the American Statistical
  Association] {10.1080/01621459.1995.10476572}, 90, 773

\bibitem[\protect\citeauthoryear{Khan, Husa, Hannam, Ohme, P\"urrer, Forteza
  \& Boh\'e}{Khan et~al.}{2016}]{PhysRevD.93.044007}
Khan S.,  Husa S.,  Hannam M.,  Ohme F.,  P\"urrer M.,  Forteza X.~J.,   Boh\'e
  A.,  2016, \mn@doi [Phys. Rev. D] {10.1103/PhysRevD.93.044007}, 93, 044007

\bibitem[\protect\citeauthoryear{Li et~al.,}{Li
  et~al.}{2012}]{PhysRevD.85.082003}
Li T. G.~F.,  et~al., 2012, \mn@doi [Phys. Rev. D]
  {10.1103/PhysRevD.85.082003}, 85, 082003

\bibitem[\protect\citeauthoryear{{Lodewyckx}, {Kim}, {Lee}, {Tuerlinckx},
  {Kuppens}  \& {Wagenmakers}}{{Lodewyckx} et~al.}{2011}]{Lodewyckx2011}
{Lodewyckx} T.,  {Kim} W.,  {Lee} M.~D.,  {Tuerlinckx} F.,  {Kuppens} P.,
  {Wagenmakers} E.-J.,  2011, J. Mathematical Psychology, 55, 331

\bibitem[\protect\citeauthoryear{Lommen}{Lommen}{2015}]{0034-4885-78-12-124901}
Lommen A.~N.,  2015, Reports on Progress in Physics, 78, 124901

\bibitem[\protect\citeauthoryear{Marin \& Robert}{Marin \&
  Robert}{2010}]{marin2010}
Marin J.-M.,  Robert C.~P.,  2010, \mn@doi [Electron. J. Statist.]
  {10.1214/10-EJS564}, 4, 643

\bibitem[\protect\citeauthoryear{Meidam et~al.,}{Meidam
  et~al.}{2018}]{PhysRevD.97.044033}
Meidam J.,  et~al., 2018, \mn@doi [Phys. Rev. D] {10.1103/PhysRevD.97.044033},
  97, 044033

\bibitem[\protect\citeauthoryear{Meng \& Wong}{Meng \&
  Wong}{1996}]{10.2307/24306045}
Meng X.-L.,  Wong W.~H.,  1996, Statistica Sinica, 6, 831

\bibitem[\protect\citeauthoryear{{Misner}, {Thorne}  \& {Wheeler}}{{Misner}
  et~al.}{1973}]{1973grav.book.....M}
{Misner} C.~W.,  {Thorne} K.~S.,   {Wheeler} J.~A.,  1973, {Gravitation}

\bibitem[\protect\citeauthoryear{Moore \& Gair}{Moore \&
  Gair}{2015}]{PhysRevD.92.024039}
Moore C.~J.,  Gair J.~R.,  2015, \mn@doi [Phys. Rev. D]
  {10.1103/PhysRevD.92.024039}, 92, 024039

\bibitem[\protect\citeauthoryear{Moore, Chua  \& Gair}{Moore
  et~al.}{2017}]{0264-9381-34-19-195009}
Moore C.~J.,  Chua A. J.~K.,   Gair J.~R.,  2017, \mn@doi [Classical and
  Quantum Gravity] {10.1088/1361-6382/aa85fa}, 34, 195009

\bibitem[\protect\citeauthoryear{Neal}{Neal}{2001}]{neal2001annealed}
Neal R.~M.,  2001, Statistics and computing, 11, 125

\bibitem[\protect\citeauthoryear{Peters \& Mathews}{Peters \&
  Mathews}{1963}]{PhysRev.131.435}
Peters P.~C.,  Mathews J.,  1963, \mn@doi [Phys. Rev.]
  {10.1103/PhysRev.131.435}, 131, 435

\bibitem[\protect\citeauthoryear{Schmidt}{Schmidt}{2002}]{0264-9381-19-10-314}
Schmidt W.,  2002, Classical and Quantum Gravity, 19, 2743

\bibitem[\protect\citeauthoryear{Scott \& Berger}{Scott \&
  Berger}{2010}]{scott2010}
Scott J.~G.,  Berger J.~O.,  2010, \mn@doi [Ann. Statist.] {10.1214/10-AOS792},
  38, 2587

\bibitem[\protect\citeauthoryear{{Sisson}}{{Sisson}}{2005}]{Sisson2005}
{Sisson} S.~A.,  2005, J. American Stat. Assoc., 100, 1077

\bibitem[\protect\citeauthoryear{Skilling}{Skilling}{2004}]{Skilling2004}
Skilling J.,  2004, American Inst. Phys. Conf. Series, 119, 1211

\bibitem[\protect\citeauthoryear{Skilling}{Skilling}{2006}]{Skilling2006}
Skilling J.,  2006, Bayesian Analysis, 1, 833

\bibitem[\protect\citeauthoryear{{Tierney} \& {Kadane}}{{Tierney} \&
  {Kadane}}{1986}]{Tierney1986}
{Tierney} L.,  {Kadane} J.~B.,  1986, J. American Stat. Assoc., 81, 82

\bibitem[\protect\citeauthoryear{Veitch \& Vecchio}{Veitch \&
  Vecchio}{2010}]{PhysRevD.81.062003}
Veitch J.,  Vecchio A.,  2010, \mn@doi [Phys. Rev. D]
  {10.1103/PhysRevD.81.062003}, 81, 062003

\bibitem[\protect\citeauthoryear{Verdinelli \& Wasserman}{Verdinelli \&
  Wasserman}{1995}]{Verdinelli1995}
Verdinelli I.,  Wasserman L.,  1995, J. American Stat. Assoc., 90, 614

\bibitem[\protect\citeauthoryear{Vigeland \& Hughes}{Vigeland \&
  Hughes}{2010}]{PhysRevD.81.024030}
Vigeland S.~J.,  Hughes S.~A.,  2010, \mn@doi [Phys. Rev. D]
  {10.1103/PhysRevD.81.024030}, 81, 024030

\bibitem[\protect\citeauthoryear{Vigeland, Yunes  \& Stein}{Vigeland
  et~al.}{2011}]{PhysRevD.83.104027}
Vigeland S.,  Yunes N.,   Stein L.~C.,  2011, \mn@doi [Phys. Rev. D]
  {10.1103/PhysRevD.83.104027}, 83, 104027

\bibitem[\protect\citeauthoryear{Villa \& Walker}{Villa \&
  Walker}{2015}]{SJOS:SJOS12145}
Villa C.,  Walker S.,  2015, \mn@doi [Scandinavian Journal of Statistics]
  {10.1111/sjos.12145}, 42, 947

\bibitem[\protect\citeauthoryear{Wetzels, Grasman  \& Wagenmakers}{Wetzels
  et~al.}{2010}]{wetzels2010encompassing}
Wetzels R.,  Grasman R.~P.,   Wagenmakers E.-J.,  2010, Computational
  Statistics \& Data Analysis, 54, 2094

\bibitem[\protect\citeauthoryear{Yunes \& Pretorius}{Yunes \&
  Pretorius}{2009}]{PhysRevD.80.122003}
Yunes N.,  Pretorius F.,  2009, \mn@doi [Phys. Rev. D]
  {10.1103/PhysRevD.80.122003}, 80, 122003

\bibitem[\protect\citeauthoryear{Yunes \& Siemens}{Yunes \&
  Siemens}{2013}]{Yunes2013}
Yunes N.,  Siemens X.,  2013, \mn@doi [Living Reviews in Relativity]
  {10.12942/lrr-2013-9}, 16, 9

\bibitem[\protect\citeauthoryear{Yunes, Yagi  \& Pretorius}{Yunes
  et~al.}{2016}]{PhysRevD.94.084002}
Yunes N.,  Yagi K.,   Pretorius F.,  2016, \mn@doi [Phys. Rev. D]
  {10.1103/PhysRevD.94.084002}, 94, 084002

\makeatother
\end{thebibliography}

\end{document}